\documentclass[%
 aip,
 amsmath,amssymb,
 reprint,%
]{revtex4-1}

\usepackage{graphicx}
\usepackage{dcolumn}
\usepackage{bm}

\usepackage[utf8]{inputenc}
\usepackage[T1]{fontenc}
\usepackage{mathptmx}
\usepackage{color}

\usepackage[deletedmarkup=xout]{changes}

\definechangesauthor[name=Zoli, color=blue]{ZD}

\begin{document}

\preprint{}

\title[Self-bias voltage formation and charged particle dynamics in multi-frequency CCPs]{Self-bias voltage formation and charged particle dynamics in multi-frequency capacitively coupled plasmas}

 \author{R. U. Masheyeva}
 \altaffiliation[]{IETP, Al-Farabi Kazakh National University, 71 Al-Farabi av., Almaty 050040, Kazakhstan}
 
 \author{K. N. Dzhumagulova}
 \altaffiliation[]{IETP, Al-Farabi Kazakh National University, 71 Al-Farabi av., Almaty 050040, Kazakhstan}
 
 \author{M. Myrzaly}
 \altaffiliation[]{IETP, Al-Farabi Kazakh National University, 71 Al-Farabi av., Almaty 050040, Kazakhstan}
 
\author{J. Schulze}
 \altaffiliation[]{Department of Electrical Engineering and Information Science, Ruhr-University Bochum, D-44780,Bochum, Germany; Key Laboratory of Materials Modification by Laser, Ion and Electron Beams, School of Physics, Dalian University of Technology, China}

\author{Z. Donk\'o}
\email{donko.zoltan@wigner.hu}
 \altaffiliation[]{Wigner Research Centre for Physics, H1121 Budapest, Hungary}


\date{\today}

\begin{abstract}
In this work, we analyze the creation of the discharge asymmetry and the concomitant formation of the DC self-bias voltage in capacitively coupled radio frequency plasmas driven by multi-frequency waveforms, as a function of the electrode surface characteristics. For this latter, we consider and vary the coefficients that characterize the elastic reflection of the electrons from the surfaces and the ion-induced secondary electron yield. Our investigations are based on Particle-in-Cell/Monte Carlo Collision simulations of the plasma and on a model that aids the understanding of the computational results. Electron  reflection from the electrodes is found to affect slightly the discharge asymmetry in the presence of multi-frequency excitation, whereas secondary electrons cause distinct changes to the asymmetry of the plasma as a function of the phase angle between the harmonics of the driving voltage waveform and as a function the number of these harmonics.

\end{abstract}

\maketitle

\section{Introduction}
\label{sec:intro}

Capacitively coupled plasma sources (CCPs) driven by radio-frequency (RF) waveforms have been aiding plasma processing industry for decades. As RF current can flow through dielectric substances as well, the electrode materials are not restricted to conducting ones, which tremendously widens the range of applications. Utilising the bombardment by high energy ions, e.g., material can be removed from surfaces (plasma etching in microelectronics). At low bombarding energies, deposition from the plasma prevails (e.g. in fabrication of photovolatic devices). Other surface properties playing an important role in, e.g., biomedicine, like wetability and biocompatibility can also be changed by plasma processing.\cite{lieberman2005principles,makabe2014plasma,chabert_braithwaite_2011,Schulze_2016}.

As the efficiency and the rates of the processes at the surfaces depend on the flux and the flux-energy distribution of the impinging species (mostly ions and radicals, but also electrons in some cases) a lot of effort has been devoted to the understanding and the optimisation of these characteristics.\cite{heil2008possibility,kawamura1999ion,qin2010tailored,zhang2015control,bruneau2016controlling,kruger2019voltage,bogdanova2020virtual} The flux of the ions is mainly defined by the plasma density, whereas the flux-energy distribution is controlled by (i) the voltage drop over the sheath adjacent to the surface, (ii) the collisionality of the sheath, and (iii) the relation between the ion transit time and the period of the RF excitation.\cite{Wild1,Wild2,Donko_2012} At high pressures, the ions flying through the sheaths collide several times with the atoms/molecules of the background gas and have, consequently, a low energy upon arrival at the electrode surfaces. In contrast, at low pressures, the ions have a long free path and can gain high energies while flying through the sheaths. When the ions cross the sheaths in a fraction of the RF period, their energy is determined by the instantaneous sheath voltage. When the ion transit time is much longer than the RF period the ion energy is largely determined by the time-averaged sheath voltage. Besides the flux-energy distribution of the ions, the angular distribution of the ions at the surfaces may become very important, e.g. when high aspect ratio trenches are "milled" into semiconductor wafers. \cite{huang2019plasma,huang2020pattern,Hartmann_2020} 

During the past decades, a number of approaches have been developed to provide additional "degrees of freedom" to control ion properties at the electrodes. The first of these has been the introduction of {\it Dual-Frequency (DF) excitation}, i.e. the simultaneous application of two radio frequency waveforms to the plasma.\cite{Kitajima,Boyle_2004,booth2009dual,o2008role,boyle2004electrostatic,georgieva2004numerical,voloshin2017modeling} The DF excitation utilises the “functional separation” of these two excitation signals: the high-frequency signal is responsible for the creation of the plasma whereas the low-frequency voltage is responsible for the acceleration of the ions. This way the amplitude of the high-frequency component controls the plasma density and, consequently, the ion flux at the surfaces while the amplitude of the low frequency signal controls the energy of the ions. The functional separation is most efficient when the two excitation frequencies are significantly different, but even in this case frequency coupling effects hinder the efficient control of the ion properties for this “classical” dual-frequency excitation.\cite{Gans,turner2006collisionless}

Another major step has been the discovery of the {\it Electrical Asymmetry Effect} (EAE) that allows to make geometrically symmetric plasma sources electrically asymmetric.\cite{heil2008possibility} This is achieved by applying a base RF and its second harmonic for the excitation of the plasma, which leads to a development of a DC self-bias voltage. It was  explained by theory\cite{Czarnetzki_2009} and subsequently confirmed by both simulations\cite{Donko_2008} and experiments\cite{Schulze_2009} that the self-bias voltage can be controlled by the phase angle between the driving voltage harmonics. As the self-bias voltage influences the voltage drops over the sheaths, the ion energy can readily be controlled whereas the ion flux as shown by subsequent studies can be kept at a reasonably constant level. The EAE also develops when geometrically asymmetric discharges are driven with specific waveforms and allows controlling the discharge properties within a wide range.\cite{schulze2011making,zhang2012separate}

Studies of the EAE were also extended to a higher number of harmonics ($N>2$), various special waveforms like {\it peaks-} and {\it valleys-}waveforms,\cite{delattre2013radio} as well as {\it sawtooth}-waveforms\cite{bruneau2014ion} have been introduced and investigated both experimentally and computationally. These waveforms, known as “{\it Tailored Voltage Waveforms}” (TVW)\cite{Lafleur_2015}, have been shown to provide large flexibility for controlling charged particle dynamics, the spatio-temporal distribution of the rates of elementary processes (e.g. ionization and excitation), the electron energy distribution function, as well as the ion properties. As peaks- and valleys-waveforms have markedly different positive and negative peak amplitudes these cause an {\it amplitude asymmetry effect} in the discharge. Sawtooth-type waveforms, on the other hand, have equal positive and negative peak amplitudes, however, notably different rising and falling slopes. These result in different sheath expansion velocities and, consequently, different rates of excitation and ionization at the two sides of the discharge, which generates an asymmetry and a DC self-bias voltage, termed as the {\it slope asymmetry effect}.\cite{bruneau2015strong} 

In electrically asymmetric discharges, the DC self-bias voltage ($\eta$) has a direct effect on the ion flux-energy distribution (IFED) at the electrodes. This makes such discharges attractive for surface processing  applications.\cite{bruneau2014effect,bruneau2014growth,ries2019ion} Moreover, in the presence of $\eta \neq 0$, the IFED-s at the two electrodes will be different, which may be advantageous in applications when a high ion energy is required at one electrode, while this is to be prevented at the other electrode. The dependence of the self-bias voltage on the amplitudes and the phases of multi-frequency waveforms has thoroughly been investigated.\cite{Czarnetzki_2009,derzsi2015experimental,donko2018ion} A discharge asymmetry was also found to be induced by {\it differing materials} of the two electrodes, represented by, e.g., different electron reflection probabilities\cite{Korolov_2016sticking} or different secondary electron yields\cite{lafleur2013secondary,korolov2013influence} or the combination of these.\cite{hartmann2020charged} More recently, discharge asymmetries induced by {\it inhomogeneous magnetic fields} have also been studied.\cite{yang2017magnetical,yang2018magnetical,oberberg2018experimental,sharma2018spatial} The investigation of the nonlinear coupling of these various asymmetry effects clearly warrants further studies. Note, that most plasma reactors used in industrial applications are geometrically asymmetric. When such a discharge is driven by a multi-frequency waveform and has different electrode materials, three types of asymmetry effects are present simultaneously. If a magnetic field is applied as well, then four non-linearly coupled types of asymmetry mechanisms will be present. Moreover, many of these applications use an electronegative gas or gas mixture, in which the asymmetry effects may differ significantly\cite{schulze2011electron,brandt2019control,zhang2011numerical,skarphedinsson2020tailored,schungel2016tailored,gibson2017controlling} from those in thoroughly investigated electropositive discharges.

The formation of the DC self-bias voltage and its dependence on the properties of the applied waveform have been studied experimentally and via simulations in a number of studies. The primary computational tool for these investigations has been the Particle-in-Cell/Monte Carlo Collisions (PIC/MCC) simulation. \cite{Birdsall_2004,Verboncoeur_2005,matyash2007particle,Tskhakaya_2007} This particle based approach is fully capable to capture kinetic effects,\cite{Basti_tutorial}therefore is well suited for the description of plasma sources operated at low pressures where non-local particle transport \cite{tsendin2010nonlocal,gallagher2012nonequilibrium,fu2020similarity,Wang2020asymmetries} appears. An analytical model\cite{Czarnetzki_2009} based on the voltage balance of the discharge has aided the understanding of the observations. 

While a lot of knowledge has accumulated in the previous studies (part of which has been reviewed above) some details of the discharge dynamics and the self-bias formation in CCPs require further studies. Our aim here is to understand the effects of the number of harmonics used to construct the excitation wavefrom and to reveal how these vary as a function of the parameters of the surface processes: (i) the reflection coefficient of the electrons at the electrodes and (ii) the ion-induced secondary electron yield. Here, the same values of these parameters will be assumed for both electrodes.

Following the introduction of the physical setting considered, in Section \ref{sec:methods} the analytic model and the basics of the computational method will be described in Sections \ref{sec:model} and \ref{sec:pic}, respectively. Section \ref{sec:results} is devoted to the presentation of the results, where we provide a very detailed analysis that goes beyond the details covered in previous studies. In particular, we examine the effects of the floating sheath potentials and the finite voltage drop over the plasma bulk on the discharge asymmetry and the self-bias voltage. Subsequently, the effects of the surface processes are discussed. A brief summary is given in Section \ref{sec:summary}.

\section{Physical system and methods}
\label{sec:methods}

In this work, we consider a capacitively coupled plasma source that has two parallel, planar electrodes. The diameter of the electrodes is assumed to be much larger than the gap between them, allowing the use of a one-dimensional model. The discharge is excited by a voltage waveform \cite{derzsi2015experimental}
\begin{equation}
    \phi(t) = \sum_{k=1}^N \phi_k \cos(k \omega_1 t + \theta_k),
    \label{eq:waveform}
\end{equation}
where $\omega_1 = 2 \pi f_1$, with $f_1$ being the "base" radio frequency, $\phi_k$ and $\theta_k$, respectively, the amplitude and the phase of the $k$-th harmonic. The amplitudes of the individual harmonics are set according to  
\begin{equation}
\phi_k = \frac{2(N-k+1)}{(N+1)^2} \phi^\ast.
\label{eq:amplitudes}
\end{equation}
$\phi^\ast$ is usually called the peak-to-peak voltage of the waveform. Note, however, that this is true only if Eq.\,(\ref{eq:waveform}) generates peaks- or valleys-types waveforms. The first of these cases is realised by setting $\theta_k=0$, $\forall k$, while the second case is realised by setting $\theta_k=0$ for the odd values of $k$ and $\theta_k=180^\circ$ for even values of $k$. 

\begin{figure}
\includegraphics[width=0.45\textwidth]{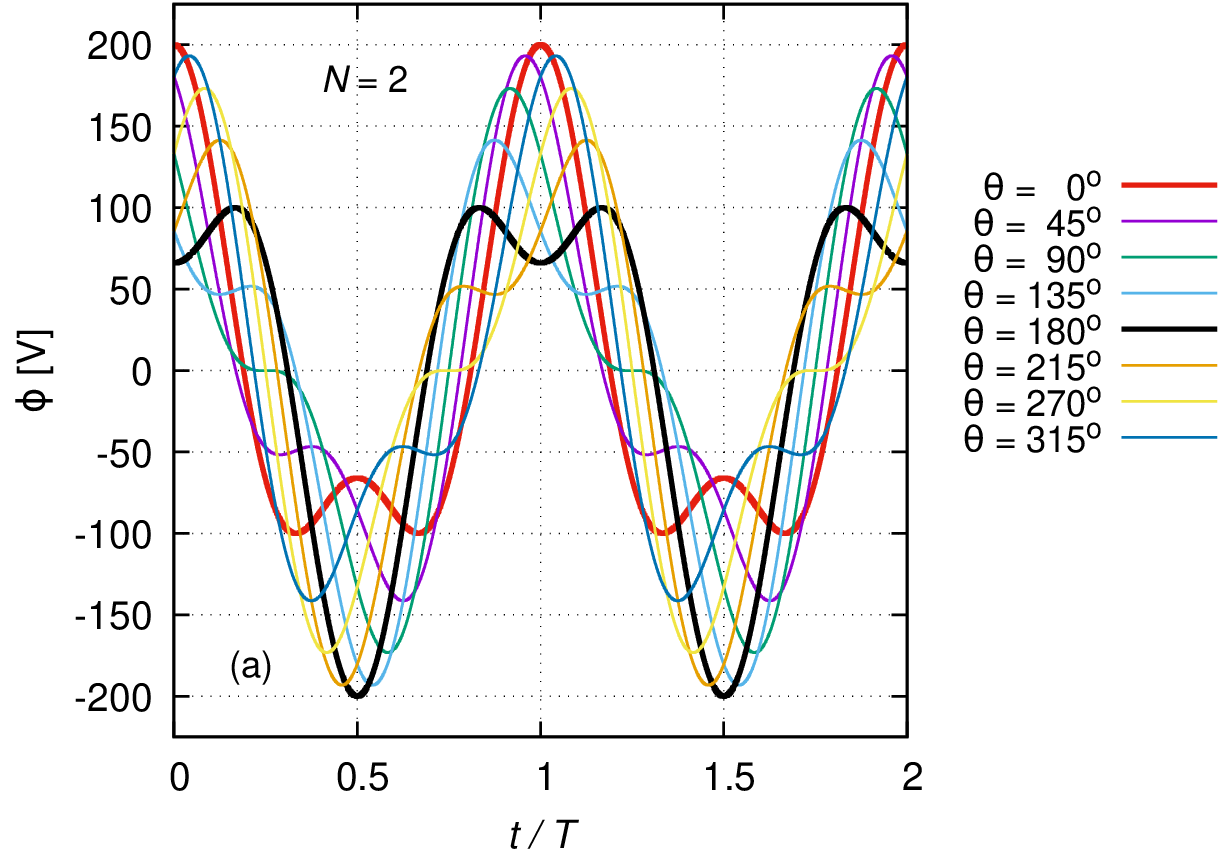}
\includegraphics[width=0.45\textwidth]{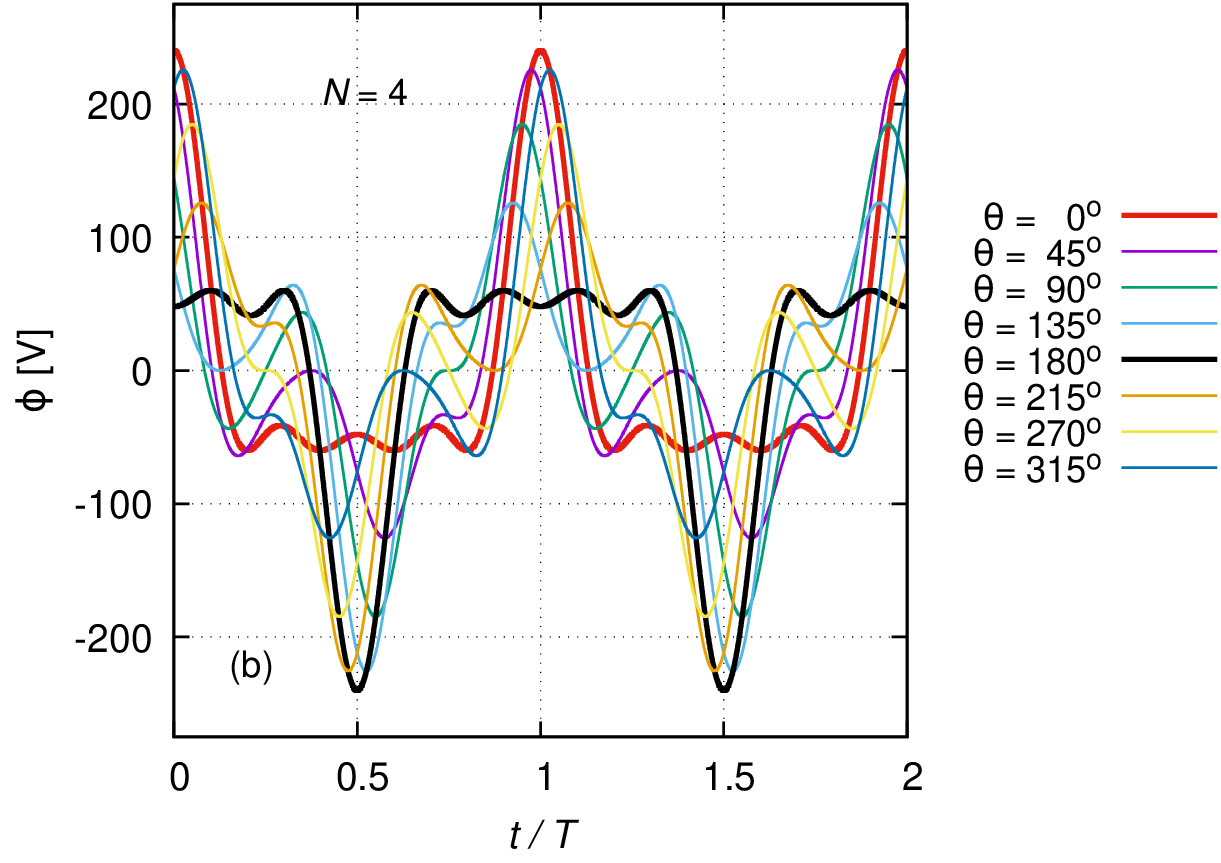}
\footnotesize 
\caption{Voltage waveforms corresponding to Eq.\,(\ref{eq:waveform}) with $\phi^\ast$ = 300\,V, for $N=2$ (a) and $N=4$ (b) harmonics, for various values of $\theta$. The peaks- and valleys-waveforms are plotted with thick lines. $T$ is the period of the fundamental frequency ($f_1$).}
\label{fig:waveforms}
\end{figure}

By keeping the phase angles of all odd harmonics $0^\circ$ and varying the common value, denoted by $\theta$, for all even harmonics, various waveforms (which include both the peaks and valleys cases) can be realised as shown in Figure \ref{fig:waveforms} for $N=2$ and $N=4$. One can note that for arbitrary values of $\theta$, the peak-to-peak amplitude of the waveform specified by (\ref{eq:waveform}) indeed varies, at $\theta = 90^\circ$, e.g., $\phi_{\rm pp} \approx 1.15 \phi^\ast$ for $N=2$ and $\phi_{\rm pp} \approx 1.23 \phi^\ast$ for $N=4$. Peaks-type waveforms ($\theta = 0^\circ$) have sharp positive peaks and nearly flat negative parts between these peaks. Correspondingly, the sheath at the powered electrode is expanded for a relatively long part of the fundamental RF period, whereas the sheath at the grounded electrode is expanded for a short time. For valleys-type waveforms ($\theta = 180^\circ$) the scenario is reversed.

\subsection{Model for the DC self-bias voltage formation}
\label{sec:model}

\begin{figure}
\includegraphics[width=0.43\textwidth]{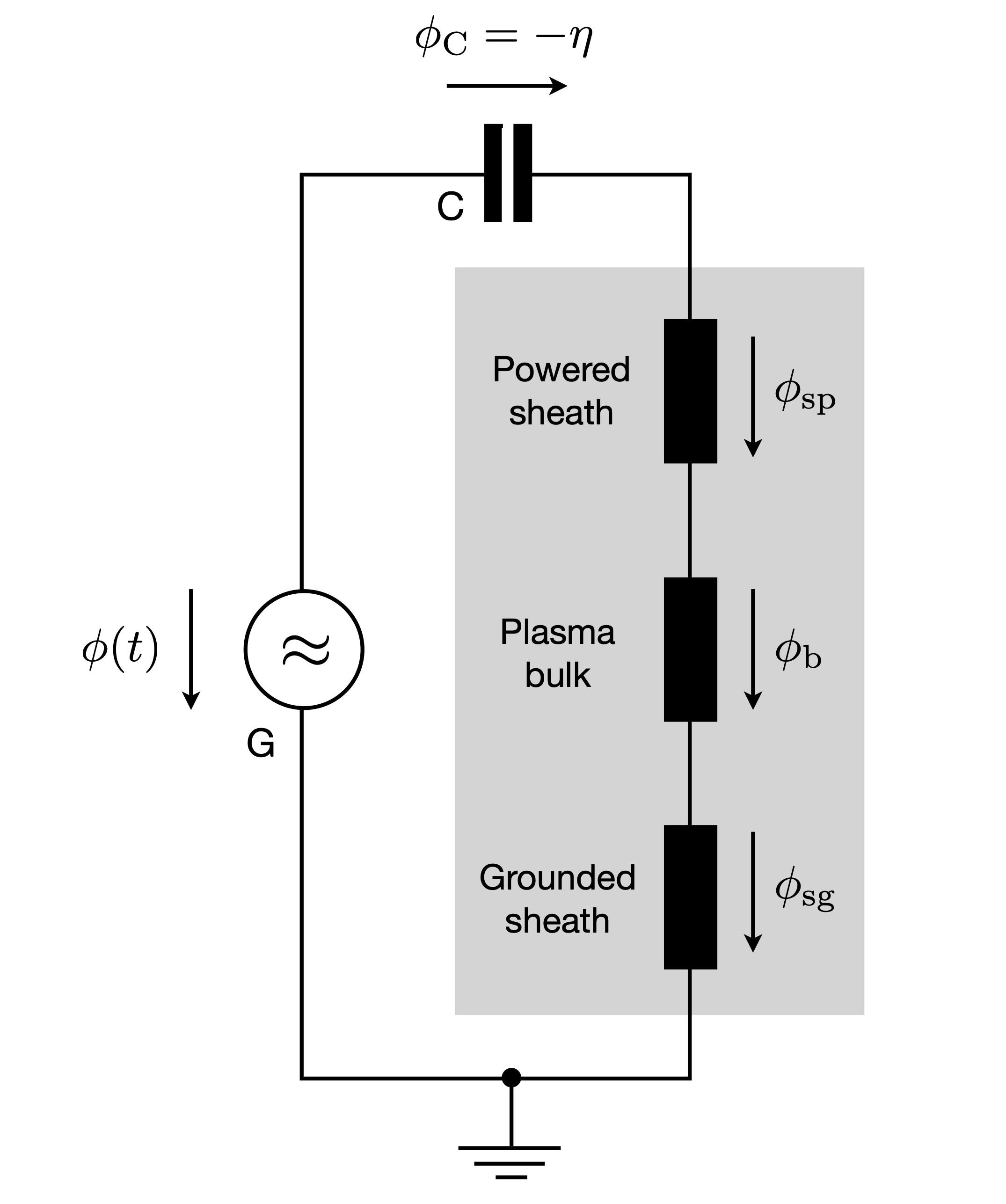}
\caption{Equivalent electrical circuit of the system investigated. The shaded area marks the plasma region, the external circuit consists of the generator G and the coupling capacitor C.}
\label{fig:circuit}
\end{figure}

Figure \ref{fig:circuit} shows the equivalent electrical circuit consisting of the RF generator (G), the blocking capacitor (C), as well as the plasma that is represented by three circuit elements corresponding to the three regions of the discharge: the sheath at the powered side of the plasma (or "powered sheath"), the plasma bulk, and the sheath at the grounded side of the plasmas (or "grounded sheath").\cite{Czarnetzki_2009} Two of these elements, the sheaths, exhibit capacitive impedance, while the impedance of the bulk region consists of a resistive and an inductive part, originating from, respectively, the finite conductivity due to electron-atom collisions and from the inertia (finite mass) of the electrons.\cite{chabert2020foundations} 

The balance equation for the voltage components marked in Figure \ref{fig:circuit} is
\begin{equation}
\phi(t) = \phi_{\rm C} + \phi_{\rm sp} + \phi_{\rm b} + \phi_{\rm sg}.
\label{eq:circuit1}
\end{equation}
Here, $\phi_{\rm C}$ is the DC voltage drop over the blocking capacitor, we assume the AC voltage drop over this element is negligible due to its high capacitance. In this case, the DC voltage drop $\phi_{\rm C}$ is the opposite of the DC self-bias voltage $\eta$ that develops over the plasma due to the EAE.\cite{Czarnetzki_2009} As a consequence, Eq.\,(\ref{eq:circuit1}) can be rewritten as
\begin{equation}
\phi(t) + \eta = \phi_{\rm sp} + \phi_{\rm b} + \phi_{\rm sg}.
\label{eq:circuit2}
\end{equation}

The model of the EAE, which assumes that (i) the sheath are fully collapsed at one side of the plasma at times of the extrema of the applied voltage waveforms and that (ii) there is no voltage drop over the bulk region of the plasma, predicts the dc self-bias voltage, based on the voltage balance of the circuit, to be
\begin{equation}
    \eta = - \frac{\phi_{\rm max}+ \varepsilon \phi_{\rm min}}{1+\varepsilon},
    \label{eq:bias-simple}
\end{equation}
where $\phi_{\rm max}$ and $\phi_{\rm min}$ are, respectively, the maximum and the minimum of the applied voltage waveform, $\phi(t)$. The more general expression, which considers the nonzero sheath voltages upon sheath collapse (i.e. the floating potentials) and the finite voltage drop over the plasma bulk,\cite{schungel2016tailored} is
\begin{equation}
    \eta =  \underbrace{- \frac{\phi_{\rm max}+ \varepsilon \phi_{\rm min}}{1+\varepsilon}}_{\eta_{\rm w}}  +
    \underbrace{\frac{\phi_{\rm sp}^{\rm f}+ \varepsilon \phi_{\rm sg}^{\rm f}}{1+\varepsilon}}_{\eta_{\rm f}} 
    +
    \underbrace{ \frac{\phi_{\rm max}^{\rm b}+ \varepsilon \phi_{\rm min}^{\rm b}}{1+\varepsilon}}_{\eta_{\rm b}}. 
    \label{eq:bias-precise}
\end{equation}
Here, we already introduced notations for the contributions of different origin: due to the waveform, $\eta_{\rm w}$, to the floating potentials, $\eta_{\rm f}$, and to the bulk voltage drop, $\eta_{\rm b}$.

In the above expressions, $\varepsilon$ is the symmetry parameter, which is the magnitude of the ratio of the peak values of the sheath voltages at both sides of the plasma \cite{Czarnetzki_2009}:
\begin{equation}
    \varepsilon = \Bigg| \frac{\widehat{\phi}_{\rm sg}}{\widehat{\phi}_{\rm sp}} \Bigg|.
    \label{eq:epsilon_original}
\end{equation}
Calculations, which are not repeated here, express the extrema of the sheath voltages as\cite{heil2008possibility}
\begin{eqnarray}
\widehat{\phi}_{\rm sp} = -\frac{1}{2e\varepsilon_0}\biggl( \frac{Q_{\rm mp}}{A_{\rm p}} \biggr)^2 \frac{I_{\rm sp}}{\overline{n}_{\rm sp}},
\label{eq:phisp}\\
\widehat{\phi}_{\rm sg} = \frac{1}{2e\varepsilon_0}\biggl( \frac{Q_{\rm mg}}{A_{\rm g}} \biggr)^2 \frac{I_{\rm sg}}{\overline{n}_{\rm sg}}.
\label{eq:phisg}
\end{eqnarray}
Here, $e$ is the elementary charge, $\varepsilon_0$ the permittivity of free space, $Q_{\rm mp/mg}$ the maximum charges within the sheaths, $A_{\rm p/g}$ the surfaces, $I_{\rm sp/sg}$ the sheath integrals, and $\overline{n}_{\rm sp/sg}$ the mean  charged particle densities within the sheaths at the powered (p) and grounded sides (g) of the system. Note, that upon the original derivation of these expressions,\cite{heil2008possibility} the electron front was assumed to exhibit a step profile at the sheath edge and the ion density profile was taken to be static within the sheath regions. Here, however, the data obtained from the PIC/MCC simulations include the slight penetration of a finite electron density into the sheaths, i.e. $Q$ and $\overline{n}$ represent, respectively, the {\it net} charge and charged particle density. 

The ratio of the sheath integrals appearing in the above expressions is customarily approximated\cite{Czarnetzki_2009} by a value of 1.0. In the symmetric system considered here, $A_{\rm p} = A_{\rm g} = A$. With these two simplifications, Eq.\,(\ref{eq:epsilon_original}) becomes:
\begin{equation}
    \varepsilon = \biggl( \frac{Q_{\rm mg}}{Q_{\rm mp}} \biggr)^2 ~
    \frac{\overline{n}_{\rm sp}}{\overline{n}_{\rm sg}}.
    \label{eq:epsilon_model}
\end{equation}
Upon the presentation of the results, the most precise of $\varepsilon$ will be computed from this equation, however, the validity of simplified approaches will also be tested, as follows:

\begin{itemize}
    \item The simplest approximation for $\varepsilon$ is represented by a value of $\varepsilon=1$, which neglects the differences between the magnitudes of the peak sheath voltages at the two electrodes. This approximation is termed as 'Model 1'.
    \item As a refinement, one may calculate $\varepsilon$ with neglecting the difference between $Q_{\rm mp}$ and $Q_{\rm mg}$ in Eq.\,(\ref{eq:epsilon_model}), i.e. taking 
    \begin{equation}
    \varepsilon = \frac{\overline{n}_{\rm sp}}{\overline{n}_{\rm sg}},
    \end{equation} which is the form that was used in the first model of the EAE.\cite{heil2008possibility} This is our 'Model 2'.
    \item Calculating $\varepsilon$ from the "full" Eq. (\ref{eq:epsilon_model}), as done in several recent studies.\cite{Schulze_2010,schungel2016tailored,brandt2019control} We refer to this as 'Model 3'.
\end{itemize}

The origin of any discharge asymmetry can also be approached from the most important elementary process in the plasmas: the ionization. This process is the primary source of the charged particles and under the conditions studied here, is driven by high-energy electrons that represent a minor fraction of the electron population.\cite{Basti_tutorial} The two basic ways of gaining enough energy for ionization (relevant at our conditions) are: acceleration of the electrons (i) near the edges of the expanding sheaths ("$\alpha$-heating")\cite{schulze2014effect} and (ii) within the sheaths in the strong electric field whenever (secondary) electrons are emitted from the electrodes, due to, e.g., ion bombardment ("$\gamma$-heating").\cite{belenguer1990transition}

The causes of asymmetry effects discussed above, like different positive vs. negative values or different rising vs. falling slopes of the driving voltage waveform, as well as different secondary electron yields at the two electrodes can also be viewed to act via establishing an imbalance of the ionization at the two sides of the plasma. A faster sheath expansion (controlled by the driving waveform), e.g., gives rise to a higher energy gain of the electrons and, generally, creates a higher charge density at the corresponding side of the plasma. In the presence of secondary electrons, the magnitude of the sheath voltages (accelerating these electrons) and the duration of the expanded phase of the sheath are the  important factors. A higher sheath voltage and/or a longer expanded phase of the sheath gives rise to higher ionization rate. All these effects can couple in a complicated nonlinear way in a CCP.


\subsection{Computational method}
\label{sec:pic}

Our numerical results are obtained from  one-dimensional (1D3V) bounded electrostatic PIC/MCC simulations. As this is a well-established method, the description of its details is omitted here, only some details specific to the current study are outlined below. More information about the approach can be found in the literature.\cite{donko2021edupic} 

Our code considers electrons and Ar$^+$ ions and follows their motion in an electric field that is defined by the potentials of the electrodes and the presence of the charged particles in the electrode gap. The powered electrode (situated at $x=0$) is at a potential $\phi(t) + \eta$, while the other electrode (situated at $x=L$) is grounded ($\phi(t)$ is defined by Eq.\,(\ref{eq:waveform})). 

The equation of motion of the charged particles is integrated using the leapfrog scheme, with a time step of $\Delta t = T/3000$. The computational grid for the potential, the electric field, and the charged particle densities (that has a spatial resolution of $\Delta x$) comprises 500 points. These parameters fulfil the relevant stability criteria of the PIC/MCC method.\cite{kim2005particle,lymberopoulos1995two} 

At the electrode surfaces, as already mentioned in Section \ref{sec:intro}, two processes are considered. (i) Ar$^+$ ions arriving at the surface induce the emission of a secondary electron with a probability that is expressed by the secondary electron yield, $\gamma$. (ii) Electrons arriving at the electrode surfaces undergo an elastic reflection event with a probability $R$ (of which the dependence on energy and angle of incidence is not taken into account). 

The DC self-bias voltage of the discharges driven by $N>1$ harmonics is determined in an iterative manner.\cite{Donko_2008} At the initialization of the simulation, $\eta = 0$\,V is set. After executing the simulation for a given number (typically 50) of RF cycles, the currents of the electrons and argon ions reaching each electrode are compared. Depending on the balance of these currents, the self-bias voltage is changed by a small quantity. This procedure is continued until $\eta$ reaches a converged value and the time-averaged charged particle currents to each of the two electrodes balance over (within the noise level).

For our studies the identification of the position of the RF sheath edge as a function of time, $s(t)$, is a crucial task. It is carried out based on computing the spatially and temporally resolved distributions of the electron and ion densities.\cite{brinkmann2007beyond} The position of the sheath sheath edge, e.g., at the electrode at $x=0$ (i.e. $s = s_{\rm p}$) can be found from
\begin{equation}
    \int_0^s n_{\rm e}(x) {\rm d}x = \int_s^h [n_{\rm i}(x)-n_{\rm e}(x)] {\rm d}x.
\end{equation}
Here, $h$ is a position where quasineutrality holds, we set this value as $h=L/2$. Solving the above equation for each time step within the RF cycle, the $s(t)$ function can be determined at both sides of the discharge. When $s_{\rm p}(t)$ and $s_{\rm g}(t)$ are known, the voltage drops over the sheaths ($\phi_{\rm sp}(t)$ and $\phi_{\rm sg}(t)$), the net space charges ($Q_{\rm p}(t)$ and $Q_{\rm g}(t)$) and mean net charged particle densities ($\overline{n}_{\rm sp}(t)$ and $\overline{n}_{\rm sg}(t)$) within the sheaths can readily be determined.

\section{Results}
\label{sec:results}

The simulations are carried out for Ar gas, at fixed values of the pressure, $p$ = 10 Pa, the electrode gap, $L$ = 2.5 cm, and the base frequency, $f_1$ = 13.56 MHz. Driving voltage waveforms specified by Eq.\,(\ref{eq:waveform}) will be used with $\phi^\ast$ = 300\,V and with up to $N=4$ harmonics, with phase angles over the whole domain of interest ($0^\circ \leq \theta \leq 360^\circ$). For the surface coefficients we adopt the following values.  (i) For the secondary electron yield we take $\gamma$ = 0, 0.2, and 0.4. For metal surfaces, $\gamma$ is expected to be rather small, $\lesssim 0.1$, while for dielectric surfaces the actual values may be well approximated by the higher $\gamma$-s adopted. (ii) For the elastic reflection coefficient of the electrons we take $R=0$ and 0.2. 

\begin{figure}[h]
\begin{center}
\includegraphics[width=0.42\textwidth]{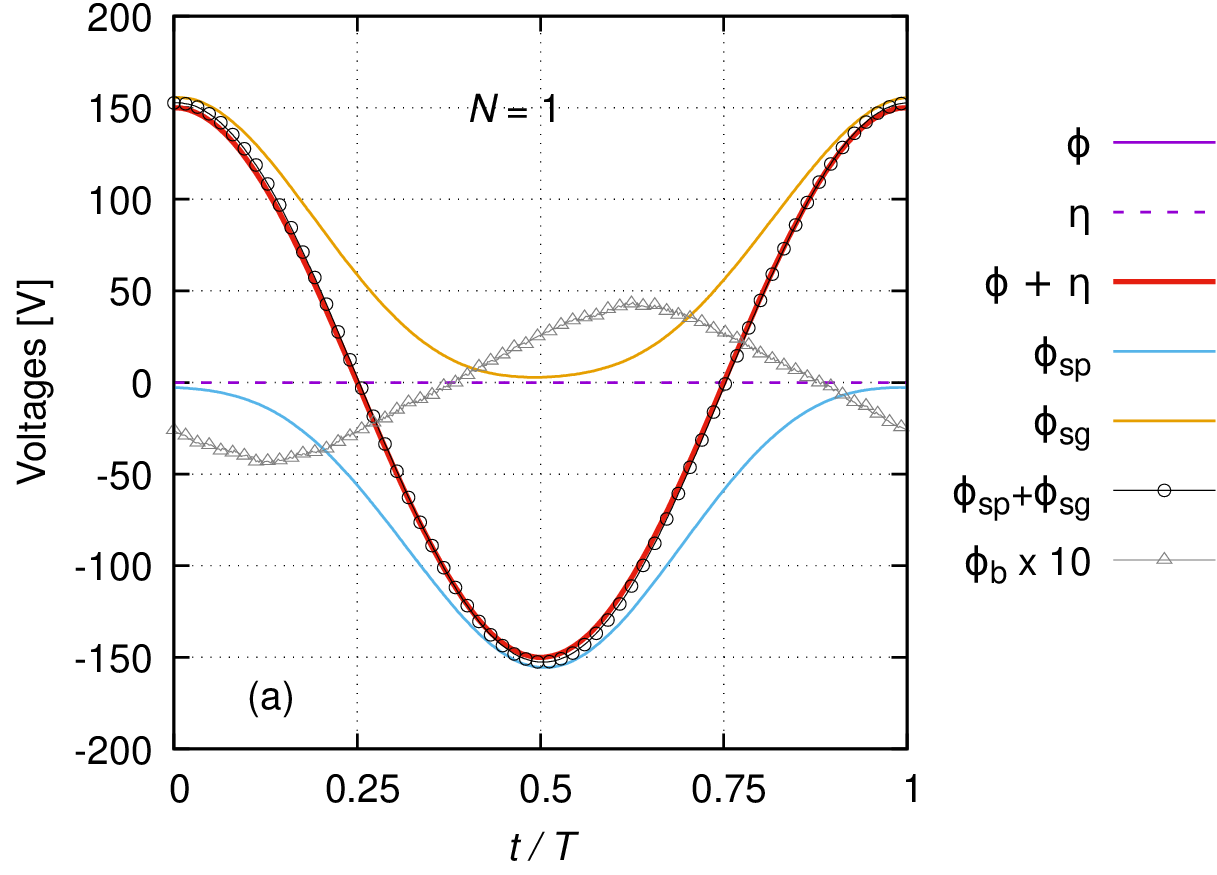}\\
\includegraphics[width=0.42\textwidth]{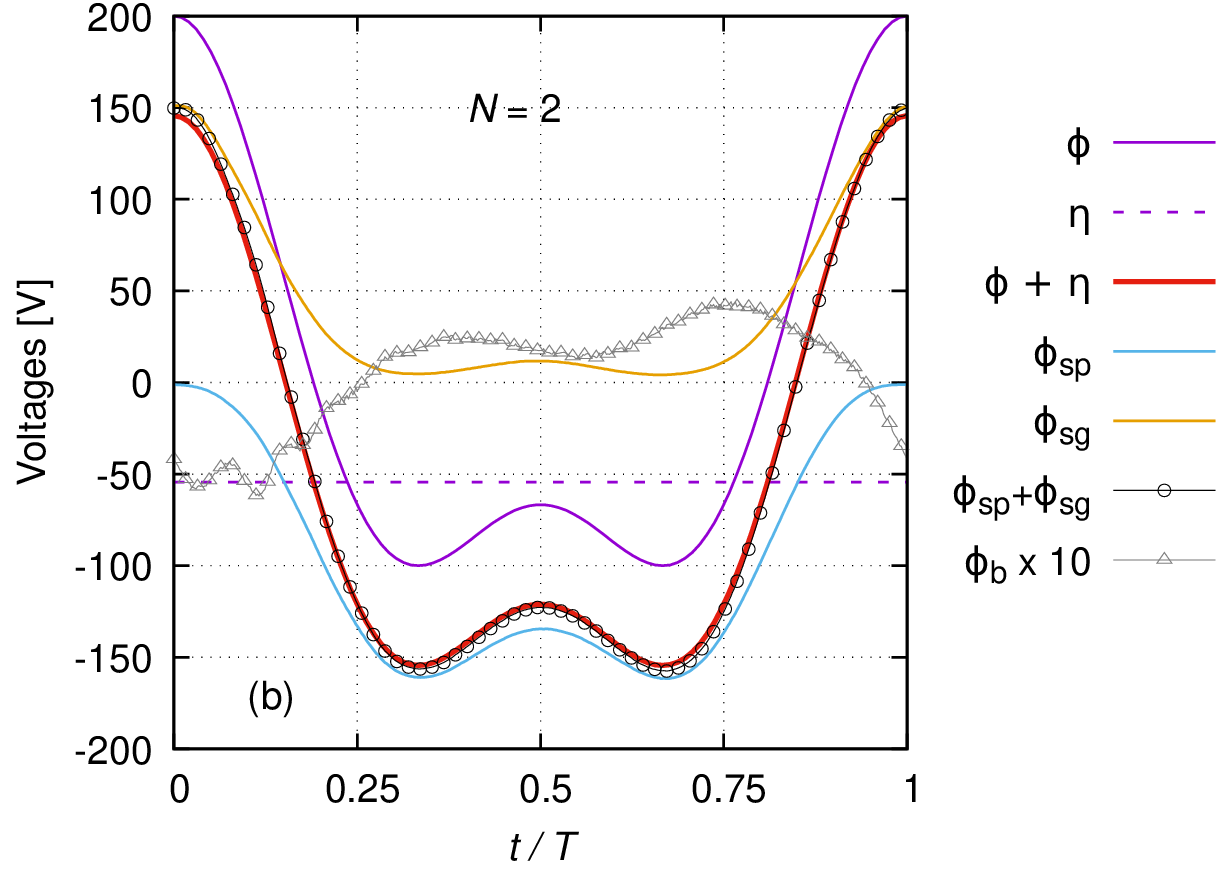}\\
\end{center}
\caption{Time dependence of the quantities involved in the voltage balance of the discharge, for single- (a) and dual-frequency (b) excitation. For $N=2$, the phase angle is $\theta=0^\circ$. Other discharge conditions: $f_1=$ 13.56\,MHz, $p=$ 10\,Pa, $L$ = 2.5\,cm, $R=$\,0, $\gamma=$\,0. The driving voltage waveform is defined by eq.\,(\ref{eq:waveform}), with amplitudes given by Eq.\,(\ref{eq:amplitudes}), $\phi^\ast$ = 300\,V. $T$ is the period of the fundamental RF frequency, $f_1$. Note, that the bulk voltage drop is multiplied by a factor of 10.}
\label{fig:voltage_example}
\end{figure}

We start the presentation of the results by discussing the temporal behavior of the various quantities that appear in the  voltage balance equation (\ref{eq:circuit2}), for the cases of single- ($N=1$) and dual-frequency ($N=2$) excitation. The $N=1$ case is displayed in Figure \ref{fig:voltage_example}(a), for the base conditions of $\phi_1$ = 150\,V, $f_1=$ 13.56\,MHz, $p=$ 10\,Pa, $L$ = 2.5\,cm, at zero values of the surface coefficients. At this electrically symmetric excitation waveform the plasma is symmetric and no self-bias voltage develops. The magnitudes of the sheath voltages ($\phi_{\rm sp}$ and $\phi_{\rm sg}$) vary with opposite phase. Note, that $\phi_{\rm sp} \leq 0$ and $\phi_{\rm sg} \geq 0$ (cf. Eqs.\,(\ref{eq:phisp}) and (\ref{eq:phisg})). The minima of $|\phi_{\rm sp}|$ and $|\phi_{\rm sg}|$ at the extrema of the applied voltage amount a few Volts. These are the so-called floating potentials, $\phi_{\rm sp}^{\rm f}$ and $\phi_{\rm sg}^{\rm f}$, respectively, at the powered and at the grounded electrodes, which limit the losses of the electrons to the electrode where the sheath momentarily collapses. The sum of the sheath voltages approximates quite well the discharge voltage (eq.\,(\ref{eq:circuit2})) as the voltage drop over the bulk of the plasmas, $\phi_{\rm b}$, amounts $\pm$\,few Volts only, due to the high conductivity of the plasma.  

The results for the $N=2$ case (with keeping all other parameters the same) are shown in Figure \ref{fig:voltage_example}(b), for the choice of $\theta=0^\circ$. The simulation reveals that for these conditions a self-bias voltage of $\eta \approx -54.4$\,V forms (indicated by the horizontal dashed line in Figure \ref{fig:voltage_example})(b)). The substantial contributions to the discharge voltage, which is the sum of $\eta$ and the generator voltage $\phi(t)$, are the sheath voltage drops, as above. A small additional contribution is provided by the nonzero voltage drop over the bulk region. 
As compared to the $N=1$ case, now the behavior of the two sheaths is quite different. Due due to the specific applied voltage waveform, the sheath at the powered electrode collapses once within the "principal" RF cycle ($T=1/f_1$) at $t/T=0$, while at the grounded side the sheath collapses twice during this period. Moreover, the sheath stays collapsed in the latter case for a longer time. As a consequence, the floating potential has a higher value of $\phi_{\rm sg}^{f} \approx 4.6$\,V as compared to the $\phi_{\rm sp}^{f} \approx -1.1$\,V found at the powered electrode. (Note, that these values are not resolved in the figure). These potentials ensure the compensation of electron and ion currents over an RF period by regulating the electron fluxes that reach the electrodes.

\begin{figure}[h]
\includegraphics[width=0.38\textwidth]{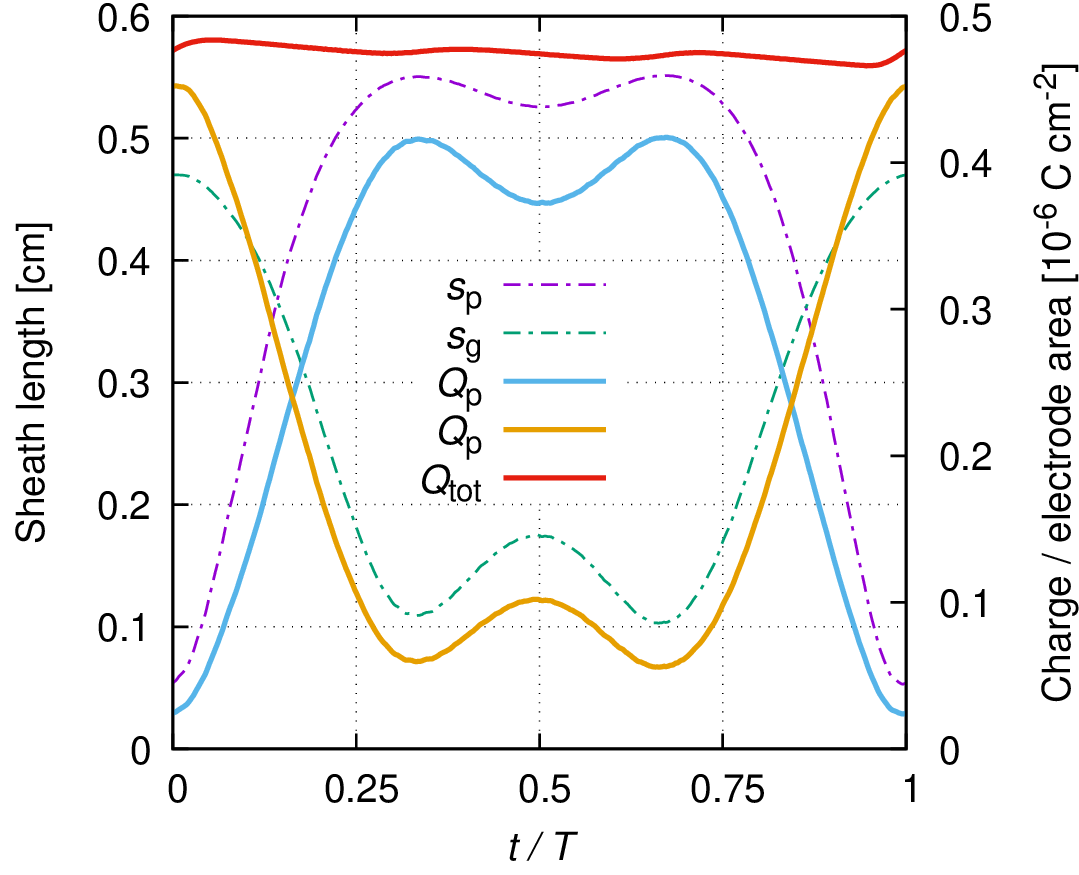}
\caption{Time dependence of the sheath lengths (chain lines, left scale) and net charge in the individual sheaths ($Q_{\rm p}$ and $Q_{\rm g}$) and their sum ($Q_{\rm tot} = Q_{\rm p}+ Q_{\rm g}$), per unit area of $A = 1$ cm$^2$ (thick solid lines, right scale), for the $N=2$, $\theta=0^\circ$ case. The other conditions are the same as in Figure \ref{fig:voltage_example}.}
\label{fig:sheath_example}
\end{figure}

The time dependence of the length of the sheaths for this case is presented in Figure \ref{fig:sheath_example}. The sheath at the powered side has a minimum length of about 0.06\,cm, whereas at the grounded side the minimum of $s_{\rm g}$ is $\approx$\,0.1\,cm. This figure also shows the net charge contained within the sheaths, for a unit electrode area of 1\,cm$^2$. The temporal change of $Q_{\rm p}$ and $Q_{\rm g}$ closely follows the variation of the length of the corresponding sheath. The sum of the charges in the two sheaths, $Q_{\rm tot}$ is almost invariant of time, as Figure \ref{fig:sheath_example} reveals. The slow drop (small negative slope) of $Q_{\rm tot}(t)$ is due to the continuous ion flux to the electrodes, while the temporary increase upon times of sheath collapses are due to the losses of the electrons to the electrodes.

\begin{figure}
\includegraphics[width=0.43\textwidth]{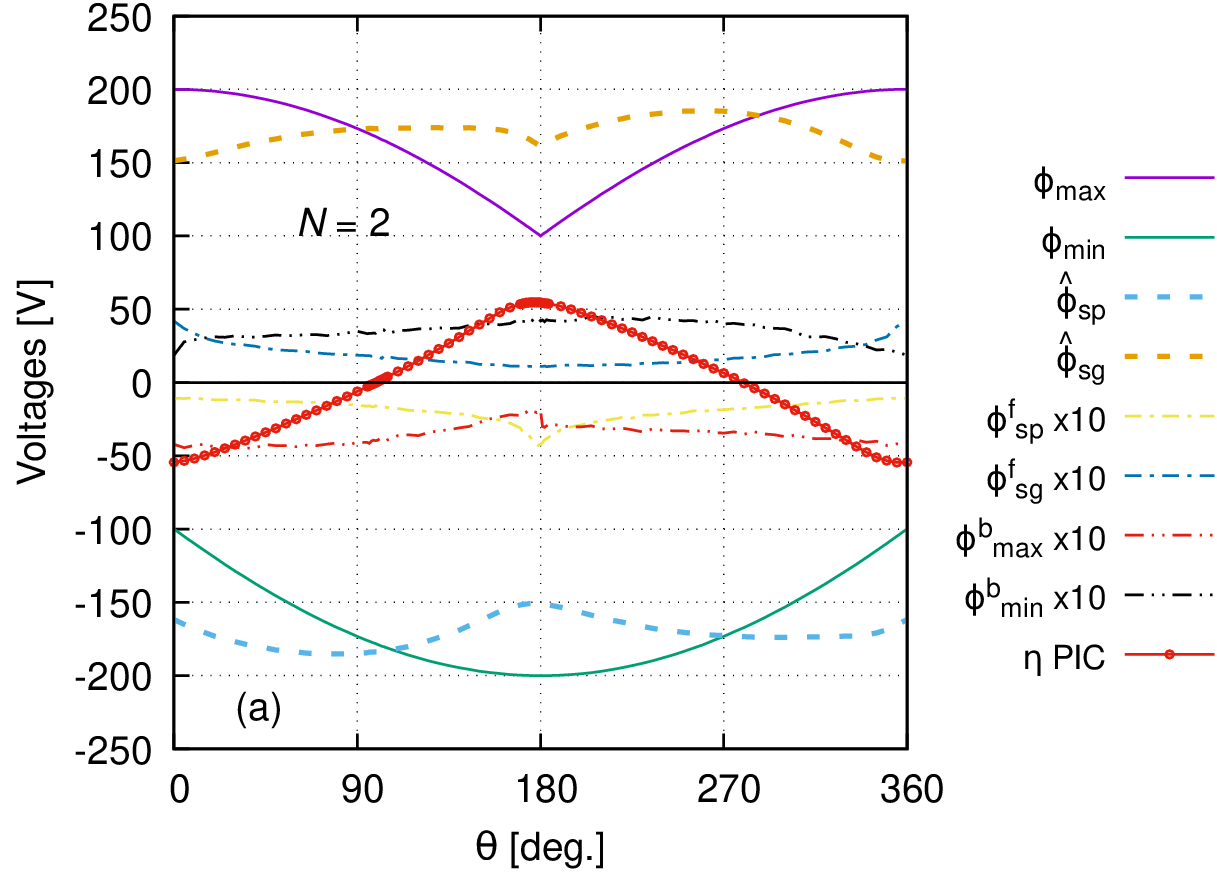}\\
\includegraphics[width=0.43\textwidth]{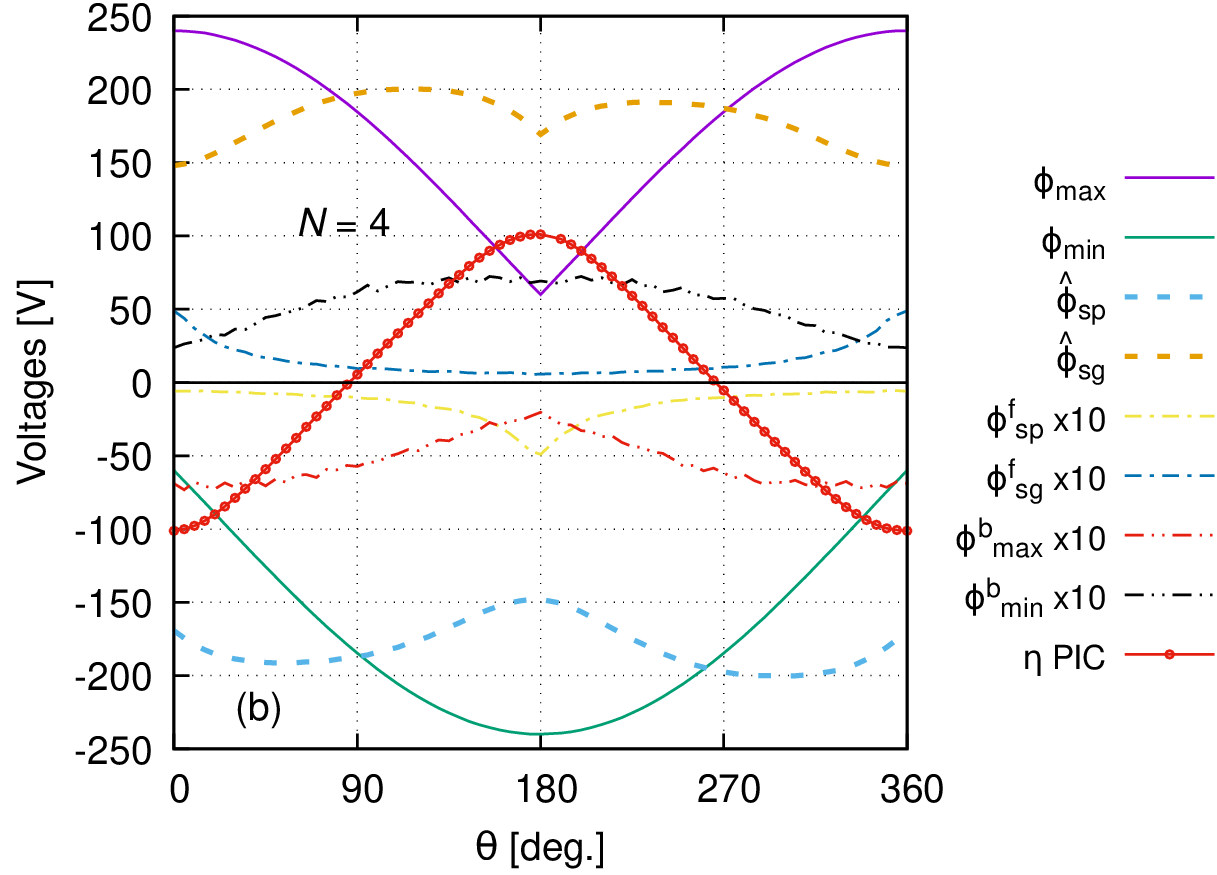}\\
 \caption{Maximum and minimum values of the applied voltage ($\phi_{\rm max}$ and $\phi_{\rm min}$), peak sheath voltages ($\widehat{\phi}_{\rm sp}$ and $\widehat{\phi}_{\rm sg}$), floating potentials ($\phi_{\rm sp}^{\rm f}$ and $\phi_{\rm sg}^{\rm f}$), bulk voltage drops at the time of the maximum and minimum of the applied voltage ($\phi_{\rm max}^{\rm b}$ and $\phi_{\rm min}^{\rm b}$), as well as the self-bias voltage $\eta$ computed from the simulation. (a) $N=2$, (b) $N=4$. Note that some quantities are multiplied by 10. Discharge conditions: Ar at $p$ = 10 Pa, $L$ = 2.5 cm, $f_1$ = 13.56 MHz, $\phi^\ast$ = 300 V, $R = 0$, $\gamma = 0$.}
\label{fig:phase1}
\end{figure}

After illustrating the time-dependent behavior of the relevant physical quantities for a few selected cases, next we address the behavior of the various voltages as a function of the phase angle $\theta$. Figure \ref{fig:phase1} shows the maximum and minimum values of the applied voltage ($\phi_{\rm max}$ and $\phi_{\rm min}$), the peak sheath voltages ($\widehat{\phi}_{\rm sp}$ and $\widehat{\phi}_{\rm sg}$), the floating potentials ($\phi_{\rm sp}^{\rm f}$ and $\phi_{\rm sg}^{\rm f}$), the bulk voltage drops at the times of the maximum and minimum of the applied voltage ($\phi_{\rm max}^{\rm b}$ and $\phi_{\rm min}^{\rm b}$), as well as the DC self-bias voltage, $\eta$, obtained directly from the simulation. (Recall that Eq.\,(\ref{eq:bias-precise}) formulates a connection between these quantities based on theory, but the comparison of these $\eta$ values with the simulation results is presented later.)

The difference between $\phi_{\rm max}$ and $\phi_{\rm min}$ equals $\phi^\ast$ (being fixed at a value of 300\,V) only at $\theta = 0^\circ$ and $180^\circ$, at other values $\phi_{\rm max}-\phi_{\rm min}>\phi^\ast$. This disparity between $\phi_{\rm max}$ and $|\phi_{\rm min}|$ is higher in the case of $N=4$, as compared to the $N=2$ case, according to the increasing asymmetry of the applied waveform (see Figure \ref{fig:waveforms}). The maxima of the magnitudes, $|\widehat{\phi}_{\rm sp}|$ and $\widehat{\phi}_{\rm sg}$ as a function of $\theta$, are on the other hand, very similar in the two cases with different number of harmonics ($N$). The floating sheath potentials show also very similar patterns in the $N=2$ and $N=4$ cases, while the voltage drop over the bulk plasma (although this is a small value), grows to almost a factor of two higher when the number of harmonics is doubled from $N=2$ to $N=4$. As to the DC self-bias voltage $\eta$, the highest values are obtained near, but not exactly at $\theta = 0^\circ$ and $180^\circ$.\cite{Schulze_2010} At $N=2$, $|\widehat{\eta}| \cong 54$\,V, while for $N=4$, $|\widehat{\eta}| \cong 100$\,V peak values are found. These values, as mentioned above, result from the simulations, where they are determined based on the balance between the electron and ion currents to the electrodes. 

\begin{figure}[h]
\includegraphics[width=0.45\textwidth]{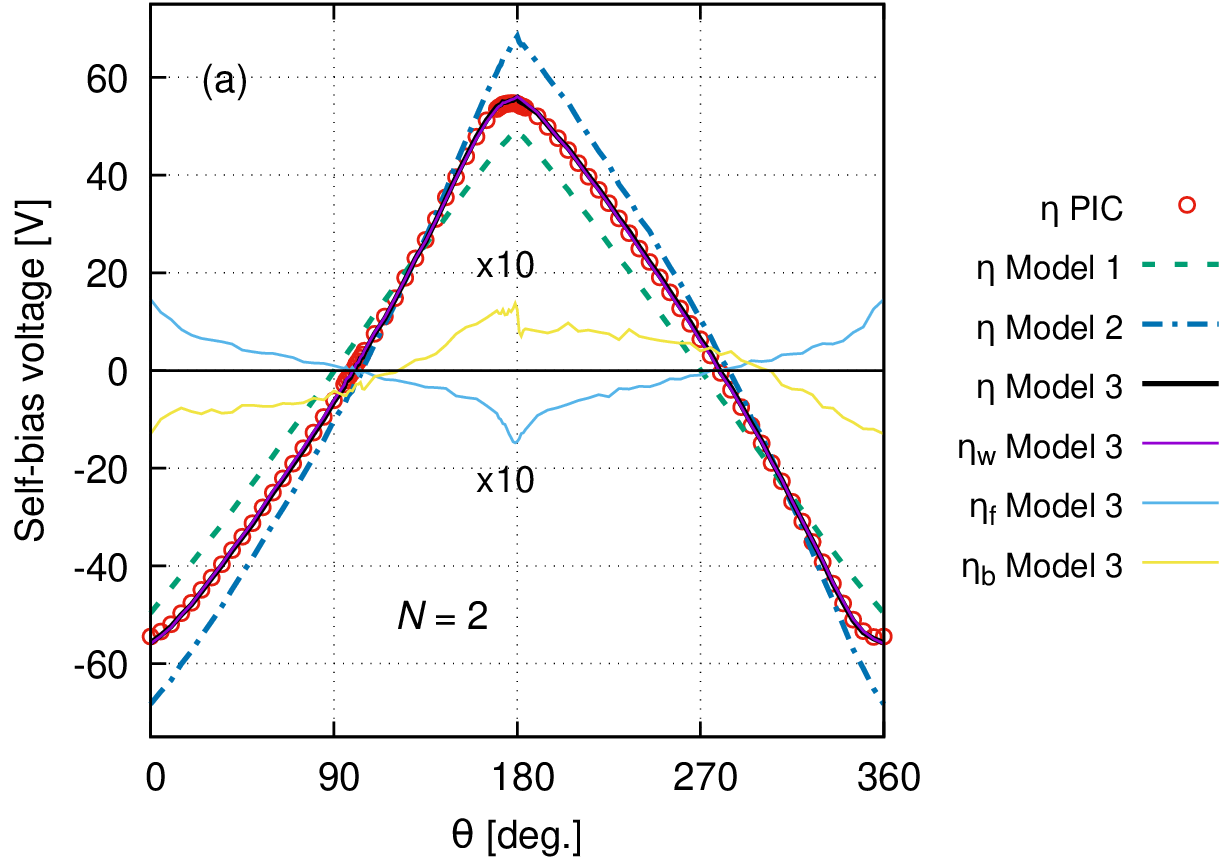}\\
\includegraphics[width=0.45\textwidth]{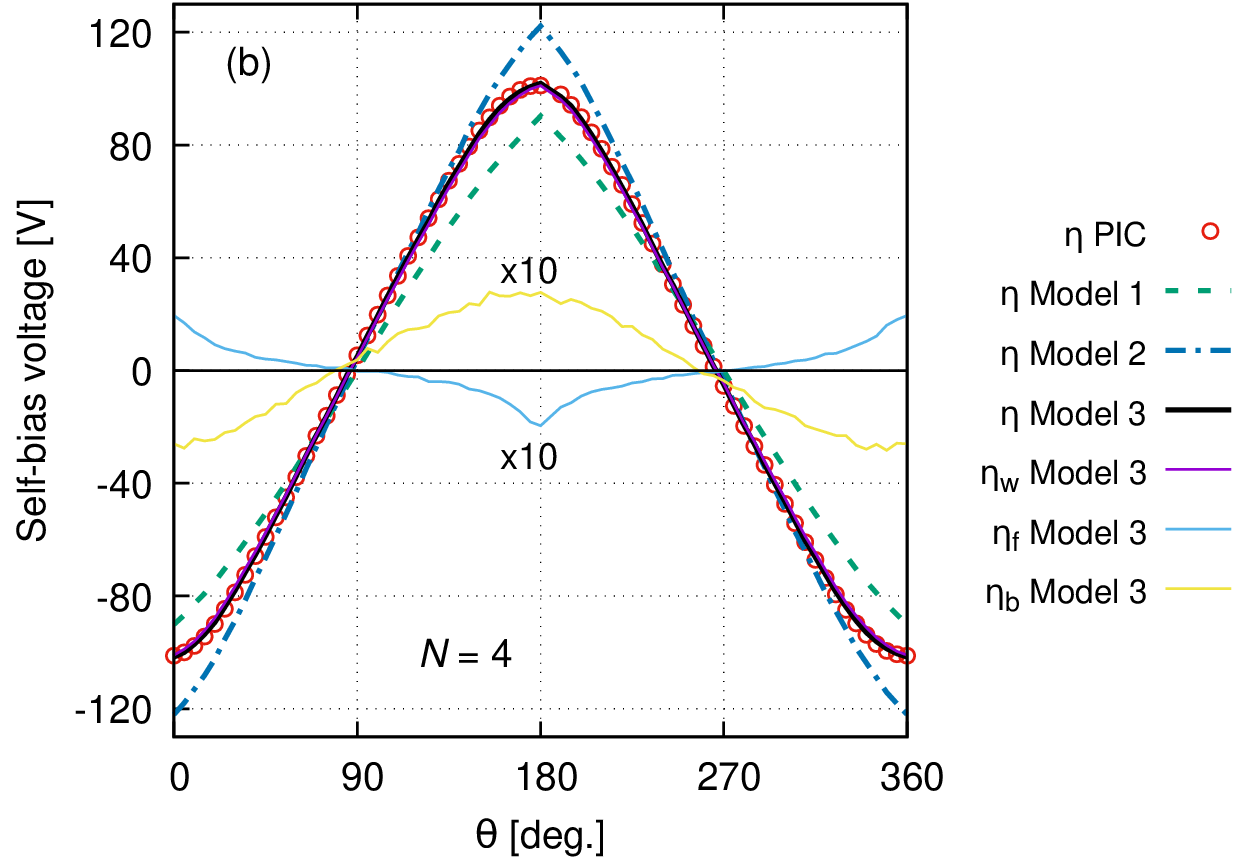}\\
 \caption{Self-bias voltage as a function of the phase angle ($\theta$) as obtained from the different models for the symmetry parameter ("$\eta$ Model 1,2,3"), as well as contributions of the terms of Eq. (\ref{eq:bias-precise}) to $\eta$ for Model 3 (Eq. (\ref{eq:epsilon_model})), in comparison with the values obtained directly from the PIC/MCC simulations ("$\eta$ PIC"). (a) $N=2$ and (b) $N=4$, for Ar at $p$ = 10 Pa, $L$ = 2.5 cm, $f_1$ = 13.56 MHz, $\phi^\ast$ = 300 V, $R = 0$, $\gamma = 0$.}
\label{fig:bias1}
\end{figure}

Next, we address the question how well the model of the EAE, outlined in Section \ref{sec:model}, reproduces these results for the self-bias voltage. For this, (i) the importance of the different terms in the expression  (\ref{eq:bias-precise}) are examined, and (ii) various approximations ("modeling levels") for the calculation of the symmetry parameter $\varepsilon$ (see the end of Sec. \ref{sec:model}) are tested. This analysis is aided by Figure \ref{fig:bias1}. The findings for $N=2$ (panel (a)) and for $N=4$ (panel (b)) are very similar, only the magnitude of $\eta$ is higher in the $N=4$ case. When Model 1 is used, i.e. $\varepsilon$ is taken to be 1.0, a triangular shape for $\eta(\theta)$ is obtained, which approximates reasonably the simulation results, although somewhat smaller $|\eta|$ values are found at the extrema of the self-bias voltage. A similar $\eta(\theta)$ dependence is found when Model 2 is adopted, however, the peak amplitudes of $\eta$ obtained this way are higher than those predicted by the PIC/MCC simulations. Finally, Model 3 provides a very good agreement with the simulation results. 

For this latter, the different contributions to the self-bias voltage, as specified in Eq. (\ref{eq:bias-precise}) are also displayed in Figure \ref{fig:bias1}. The contributions of the floating potentials and the bulk voltage drop prove to be small and acting against each other over the whole range of the phase angle $\theta$. The dominant term, $\eta_{\rm w}$, is thus hardy distinguishable from the sum of the three terms that yields $\eta$. The above observations make us conclude that consideration of the first term only in Eq. (\ref{eq:bias-precise}) is sufficient, however, for the calculation of $\varepsilon$ the more precise form of Eq. (\ref{eq:epsilon_model}) is required. 

\begin{figure}[h]
\includegraphics[width=0.45\textwidth]{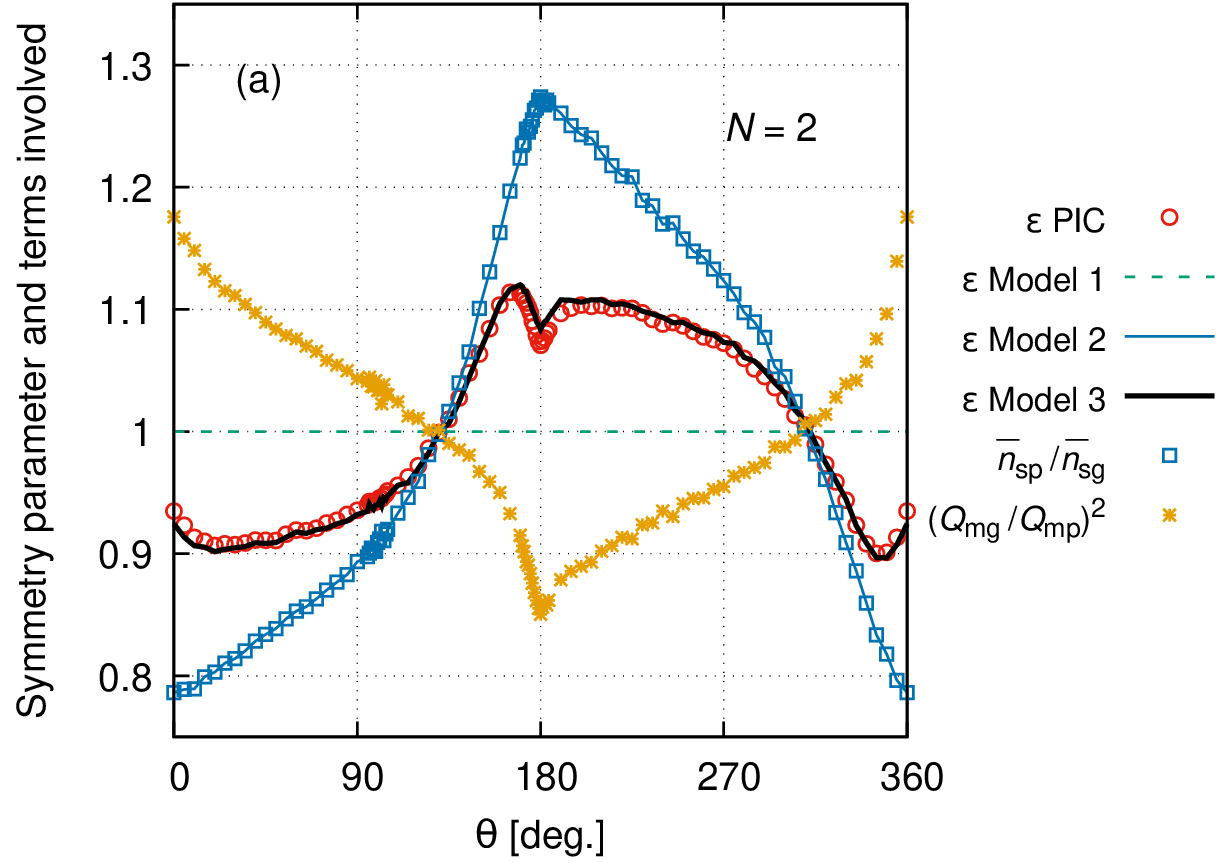}\\
\includegraphics[width=0.45\textwidth]{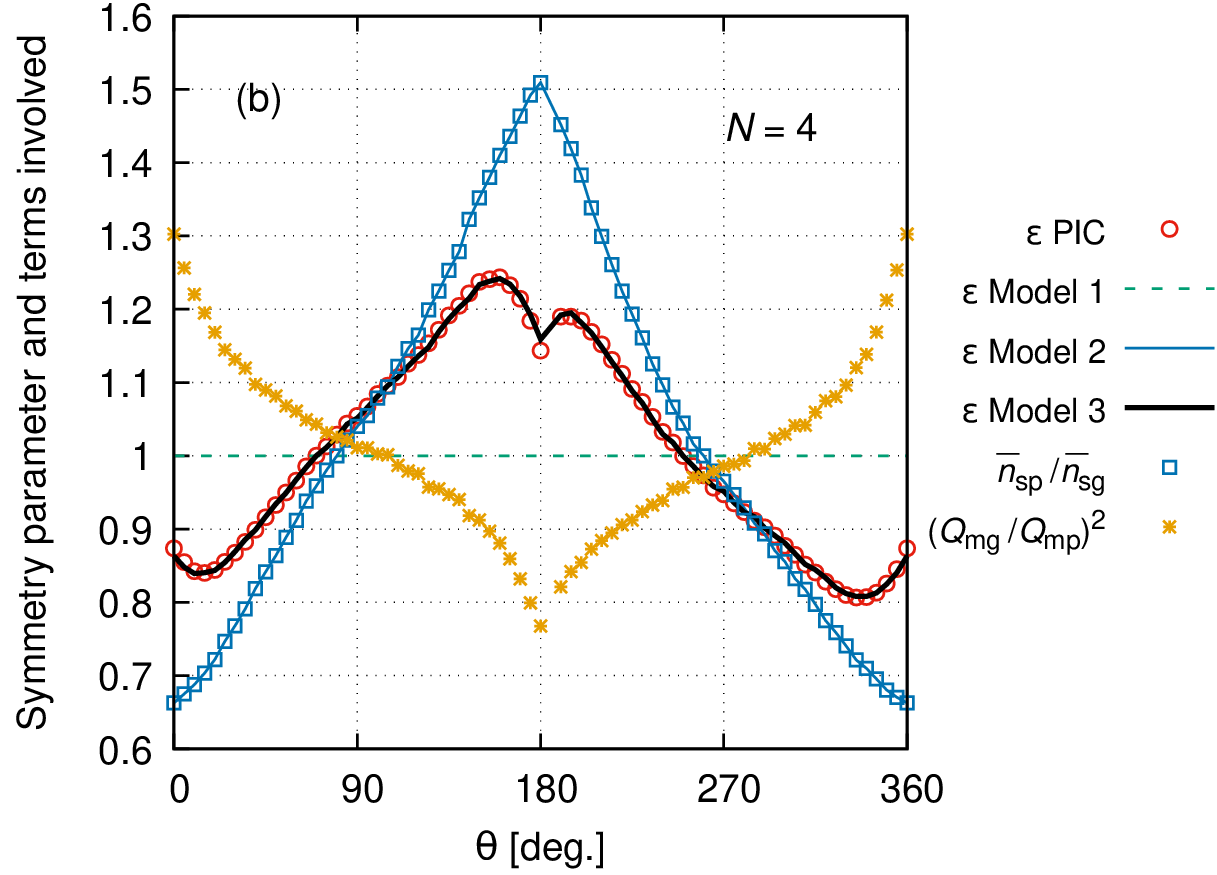}\\
 \caption{Symmetry parameter $\varepsilon$ obtained from the PIC/MCC simulation and from the different models, as well as the terms involved in the calculation of $\varepsilon$ from Eq. (\ref{eq:epsilon_model}). Note that "$\varepsilon$ Model 2" is equivalent to $\overline{n}_{\rm sp}/\overline{n}_{\rm sg}$, and that "$\varepsilon$ Model 3" is computed as $(Q_{\rm mg}/Q_{\rm mp})^2 (\overline{n}_{\rm sp}/\overline{n}_{\rm sg})$ (see Eq. (\ref{eq:epsilon_model})). Discharge conditions: Ar at $p$ = 10 Pa, $L$ = 2.5 cm, $f_1$ = 13.56 MHz, $V_{\rm pp}$ = 300 V.}
\label{fig:epsilon1}
\end{figure}

At this point it is useful to analyze the $\varepsilon$ values obtained from the different models, together with its values taken directly from the simulation (via Eq. (\ref{eq:epsilon_original})). The corresponding data are shown in Figure \ref{fig:epsilon1} as a function of the phase angle $\theta$ (panel (a) for $N=2$ and (b) for $N=4$). Additionally, the terms involved in the calculation of $\varepsilon$ via Eq. (\ref{eq:epsilon_model}) are also displayed in Figure \ref{fig:epsilon1}. As in the case of the self-bias voltage (see Figure \ref{fig:bias1}), the most accurate model (i.e. Model 3) reproduces very well the symmetry parameter as a function of $\theta$ resulting from the simulation. The $\varepsilon =1$ approximation of Model 1 is clearly a bad choice, as $\varepsilon$ varies with $\theta$ about $\pm 10\%$ in the $N=2$ case and more than $\pm 20\%$ in the case of $N=4$. Model 2 (which considers only the difference of the mean charge densities within the sheaths), on the other hand, largely (by a factor of $\approx 2$) overestimates the variation of $\varepsilon$ with $\theta$. These findings confirm that the charge dynamics, represented by the term $(Q_{\rm mg}/Q_{\rm mp})^2$ plays an important role.\cite{Schulze_2010} The inclusion of this term in the calculation of $\varepsilon$ ("$\varepsilon$ Model 3" in Figure \ref{fig:epsilon1}) yields a very good agreement with the simulation data ("$\varepsilon$ PIC").

\begin{figure}[h]
~~~~~~~~~~~~~~\includegraphics[width=0.44\textwidth]{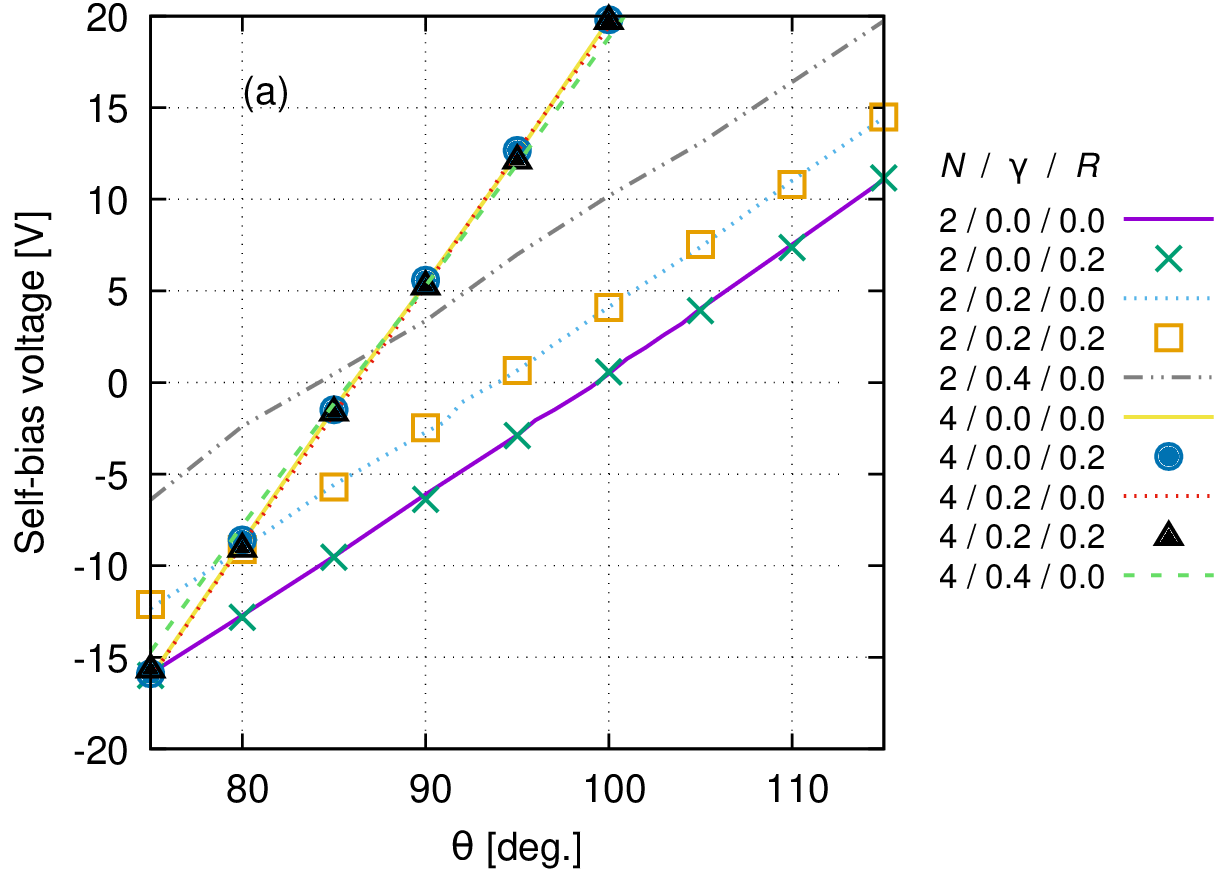}
\includegraphics[width=0.38\textwidth]{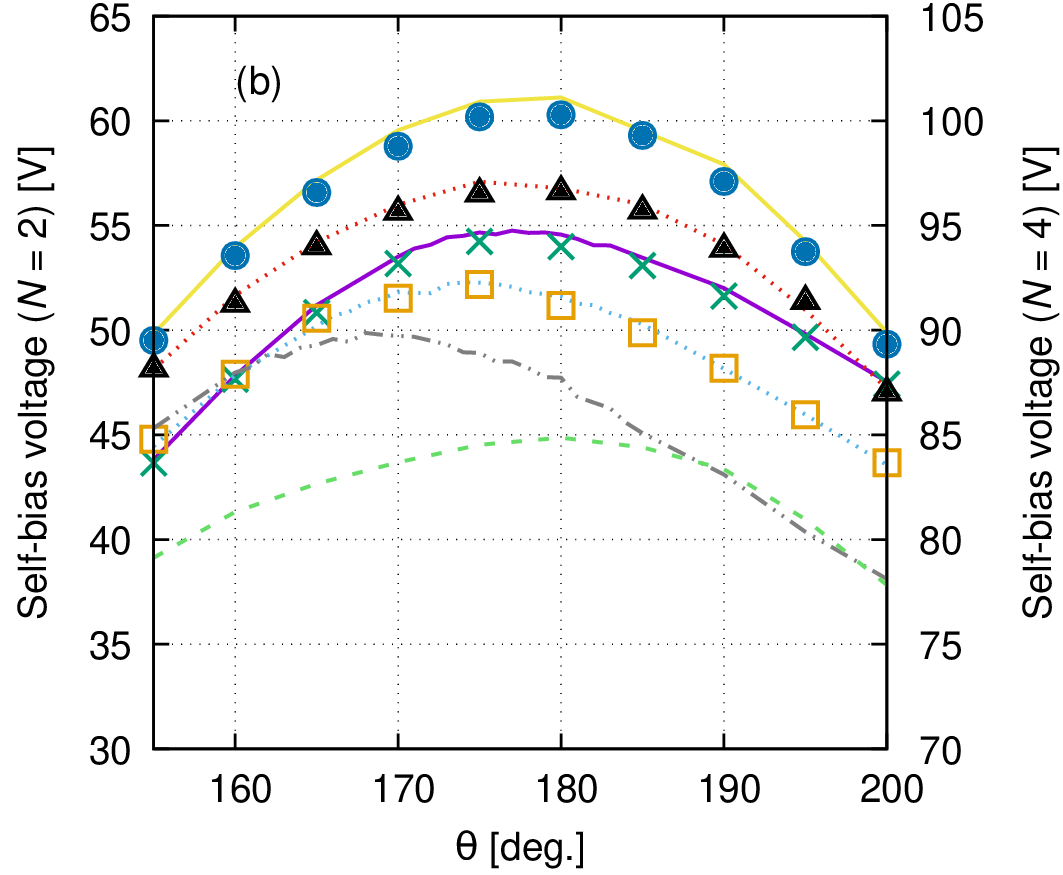}
\caption{Self-bias voltage obtained from the PIC/MCC simulations in the vicinity of  $\theta = 90^\circ$ (a) and $\theta = 180^\circ$ (b), for different number of harmonics ($N$), secondary electron yields ($\gamma$), and  electron reflection coefficients ($R$). The legend in (a) also holds for panel (b). In (b), the data corresponding to $N=2$ are given in the left axis, while data for $N=4$ are shown in the right axis. Discharge conditions: Ar at $p$ = 10 Pa, $L$ = 2.5 cm, $f_1$ = 13.56 MHz, $V_{\rm pp}$ = 300 V.}
\label{fig:bias_details_pic}
\end{figure}


Next, we analyze the behavior of the discharge in the vicinity of phase angles where the self-bias voltage is (i) zero and where (ii) it is maximised. The first domain corresponds to phase angles near $\theta = 90^\circ$, while for the second domain $\theta$ is around $180^\circ$. These two domains, respectively, are shown in Figures \ref{fig:bias_details_pic}(a) and (b), for various values of the number of harmonics. Here we also include data obtained with different values of the secondary electron yield and the electron reflection coefficient. We recall that  the effects of both of these parameters were already analyzed in the context of the EAE.\cite{lafleur2013secondary,korolov2013influence,Korolov_2016sticking} These studies have, however, focused on {\it establishing} an asymmetry by using {\it unequal} coefficients at the two electrodes, while here we study the effects of these parameters with {\it equal} values at both electrodes, i.e., they are not the causes of the asymmetry, but may modify it by influencing the plasma behavior. Based on Figure \ref{fig:bias_details_pic} the following observations can be made:

\begin{enumerate}
    \item The electron reflection has a marginal effect on $\eta$, at all values of $\theta$, $N$, and $\gamma$ considered. 
    \item At $\gamma=0$, the zero crossing of $\eta$ for $N=2$ occurs at $\theta>90^\circ$, while for $N=4$ it occurs at $\theta<90^\circ$.
    \item At $N=2$, the angle where $\eta$ becomes zero changes with $\gamma$, while for $N=4$ no such dependence is observed.
    \item The increase of $\gamma$ decreases the maximum self-bias voltage at both $N$ values. 
\end{enumerate}

\begin{figure}[]
\includegraphics[width=0.45\textwidth]{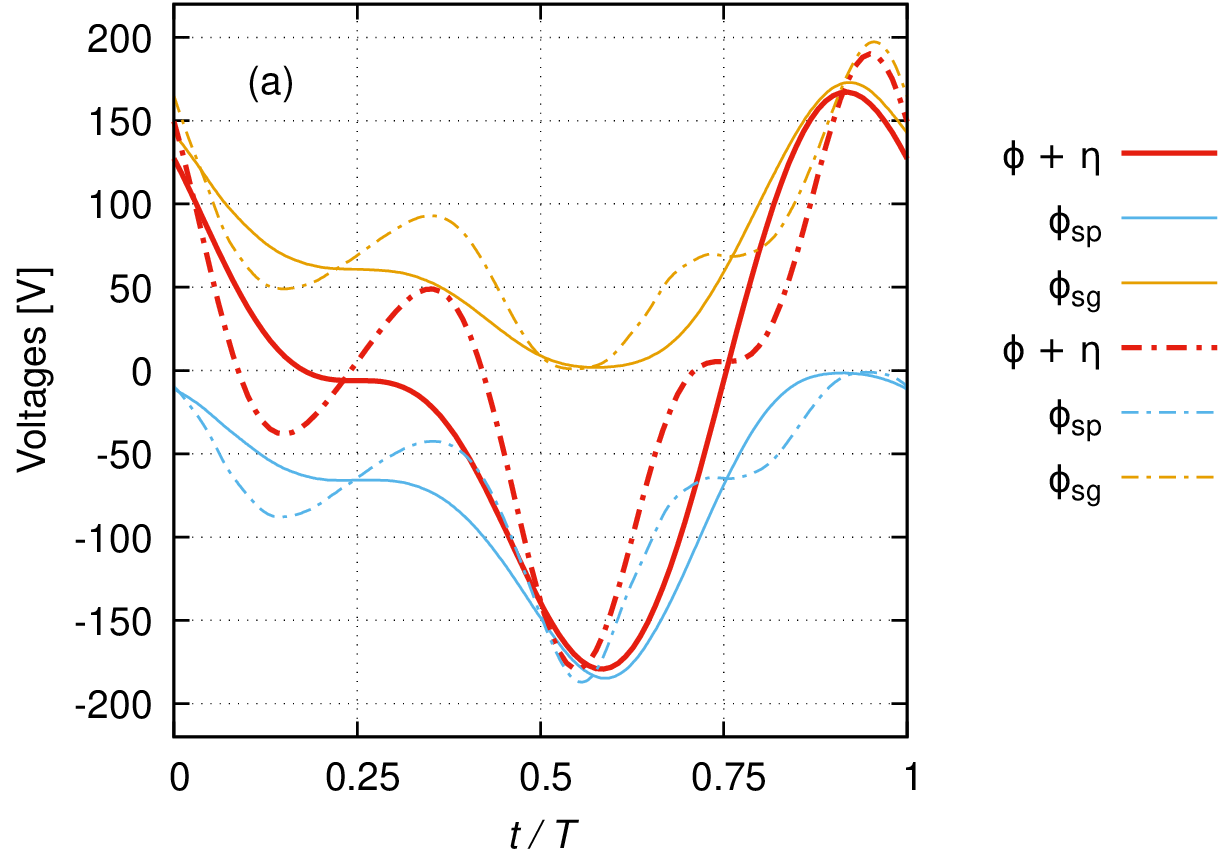}\\
~~~~~~~\includegraphics[width=0.45\textwidth]{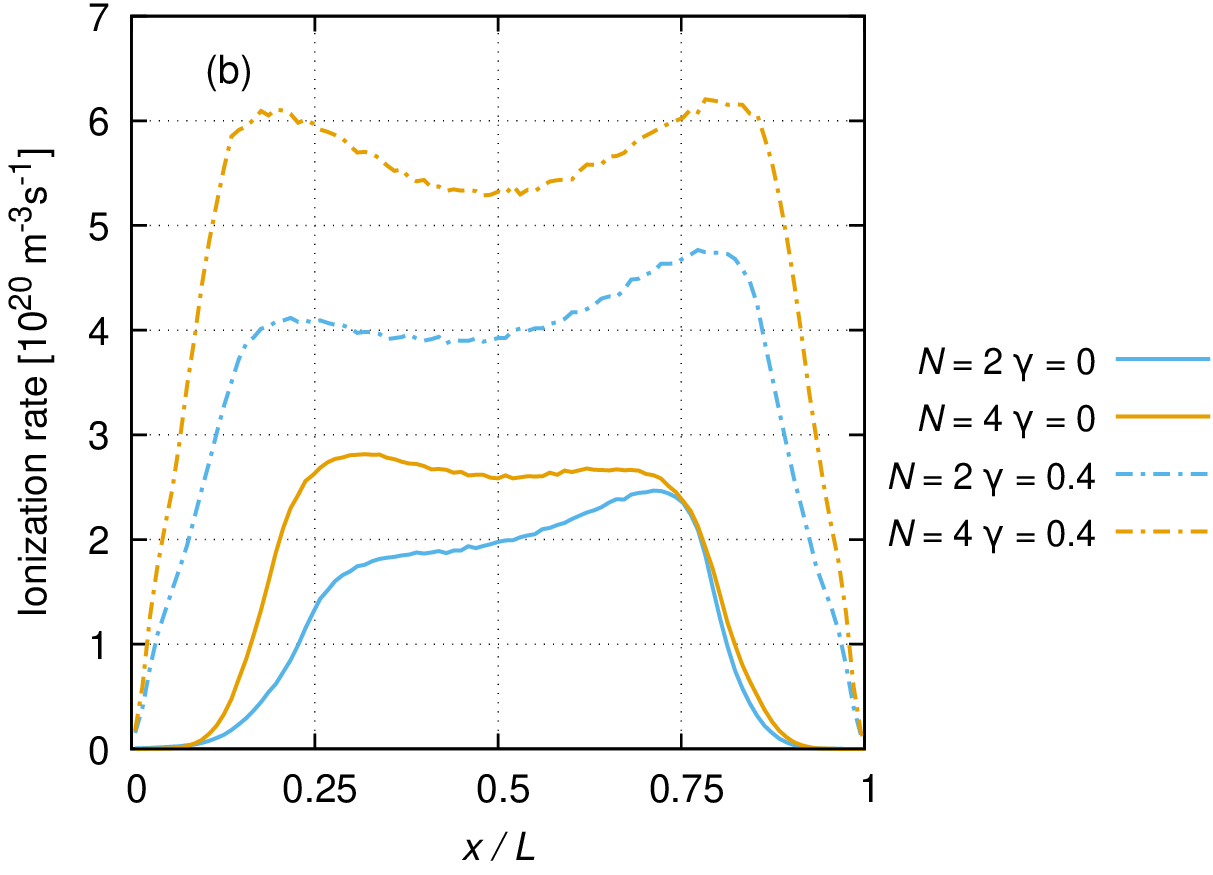}
\caption{(a) Discharge voltage ($\phi+\eta$) and sheath voltages as a function of time, at $\gamma$ = 0. Solid lines: $N=2$, chain lines: $N=4$. (b) Spatial distribution of the time-averaged ionization rate for the conditions indicated. Other conditions: Ar at $p$ = 10 Pa, $L$ = 2.5 cm, $f_1$ = 13.56 MHz, $\phi^\ast$ = 300 V, $R$ = 0. $\theta=90^\circ$ for all cases.}
\label{fig:slope}
\end{figure}

In the following, we provide explanations for these observations. The first of these can be explained by the fact that although reflected electrons generally increase the plasma density, at the relatively low value of $R$ considered here this increase is small.\cite{Korolov_2016sticking} We observe only a slight effect of electron reflection at $\gamma=0$, comparison of the pairs of data sets for $R=0$ vs. 0.2 in Figure \ref{fig:bias_details_pic}(b) for $N=2$ and 4 reveals a small decrease of $\eta$ when $R$ is increased.

Regarding the second observation, it is important to realise that at $\theta=90^\circ$ the maximum and minimum values of the driving voltage waveform have the same magnitudes ($\phi_{\rm max} = - \phi_{\rm min}= \phi_{\rm m}$), i.e. {\it no} amplitude asymmetry is present at this specific $\theta$. According to Eq. (\ref{eq:bias-simple}), any deviation of $\eta$ from zero can only be attributed to $\varepsilon \neq 1$, as $ \eta / \phi_{\rm m}= - (1-\varepsilon)/(1+\varepsilon)$. (Using Eq. (\ref{eq:bias-simple}) instead of the more precise Eq. (\ref{eq:bias-precise}) is justified by our earlier conclusion that the floating potential and the bulk voltage drop act against each other in the latter equation.) Having ruled out the effect of different (positive vs. negative) voltage amplitudes, the observed asymmetry of the discharge can only originate from the specific shapes of the driving voltage waveform. Indeed, it turns out that it is the {\it slope} of the waveform, i.e., $d\phi(t)/dt$ that is responsible for the observed effect. At $N=2$, as shown (by the red solid line) in Figure \ref{fig:slope}(a), the driving waveform has a long falling slope and a short rising slope, i.e. it resembles a sawtooth-down waveform.\cite{bruneau2015strong} This shape is expected to result in higher excitation/ionization rate at the grounded side of the discharge and this is confirmed by the computed ionization rate function shown in Figure \ref{fig:slope}(b) for $N=2$ and $\gamma=0$. The theory of sawtooth waveforms\cite{Lafleur_2015} predicts a negative self-bias voltage, in accordance with our observation in Figure \ref{fig:bias_details_pic}(a). For $N=4$, we find the fastest sheath expansion around times of $t/T \approx 0.4$ at the powered side, and the ionization source (see Figure \ref{fig:slope}(b)) exhibits a peak at this side of the discharge. This explains the observation of a small positive self-bias voltage (that differs from the $N=2$ case).


\begin{table}[h!]
\caption{Self-bias voltage ($\eta$) at $\theta=90^\circ$, at $N=2$ and 4, as a function of $\gamma$, the values of the symmetry parameter $\varepsilon$ and the terms involved in  $\varepsilon$, as well as the peak sheath voltages. \label{table:crossing90} }
\begin{tabular}{cccccccc}
\hline
$N$ & $\gamma$ & $\eta$ [V] & $\overline{n}_{\rm sp} / \overline{n}_{\rm sg}$ & $(Q_{\rm mg} / Q_{\rm mp})^2$ & $\varepsilon$ (Model 3) & $\widehat{\phi}_{\rm sg}$ [V] & $|\widehat{\phi}_{\rm sp}|$ [V]\\ 
\hline
2 & 0 & $-$6.10 & 0.893 & 1.043 & 0.932 & 172.9 & 184.8\\
2 & 0.2 & $-$2.71 & 0.920 & 1.054 & 0.969 & 175.3 & 180.7\\
2 & 0.4 & $+$3.36 & 0.967 & 1.078 & 1.042 & 180.5 & 173.6\\
\hline
4 & 0 & $+$5.40 & 1.040 & 1.011 & 1.055 & 197.3 & 187.1\\
4 & 0.2 & $+$5.21 & 1.029 & 1.023 & 1.054 & 196.4 & 186.3\\
4 & 0.4 & $+$5.24 & 1.008 & 1.043 & 1.056 & 195.1 & 184.8\\
\hline
\end{tabular}
\end{table}

The explanation of the third observation is aided by tabulated values of the DC self-bias voltage, the symmetry parameter and its relevant terms, as well as the peak sheath voltages for various values of $\gamma$, at $\theta=90^\circ$, included in Table \ref{table:crossing90}. The $\gamma=0$ case was discussed above and the origin of $\eta<0$ at $N=2$ was clarified. It is important to realise that at $\gamma>0$ the ionization balance is also influenced by secondary electrons (emitted from the electrode surfaces). The energy gain and the multiplication of these electrons, and the consequent ionization are sensitive functions of the accelerating voltage, i.e. {\it the peak sheath voltage drop}. For $N=2$, we find that $\widehat{\phi}_{\rm sg} < |\widehat{\phi}_{\rm sp}|$ at $\gamma=0$. Consequently, when secondary emission sets on at $\gamma>0$, ionization at the powered side of the discharge increases by a higher amount as compared to that at the grounded side. This is the reason why we find an increasing value of $\overline{n}_{\rm sp} / \overline{n}_{\rm sg}$ (seen in Table \ref{table:crossing90}), which, in turn results in an increase of $\varepsilon$ with $\gamma$. The other factor involved in $\varepsilon$, $(Q_{\rm mg} / Q_{\rm mp})^2$, also increases with increasing $\gamma$, further contributing to the  
change from $\varepsilon<1$ to $\varepsilon>1$ while $\gamma$ reaches  0.4. This change of $\varepsilon$ results in a switch of the sign of $\eta$. 

\begin{figure}[]
\includegraphics[width=0.48\textwidth]{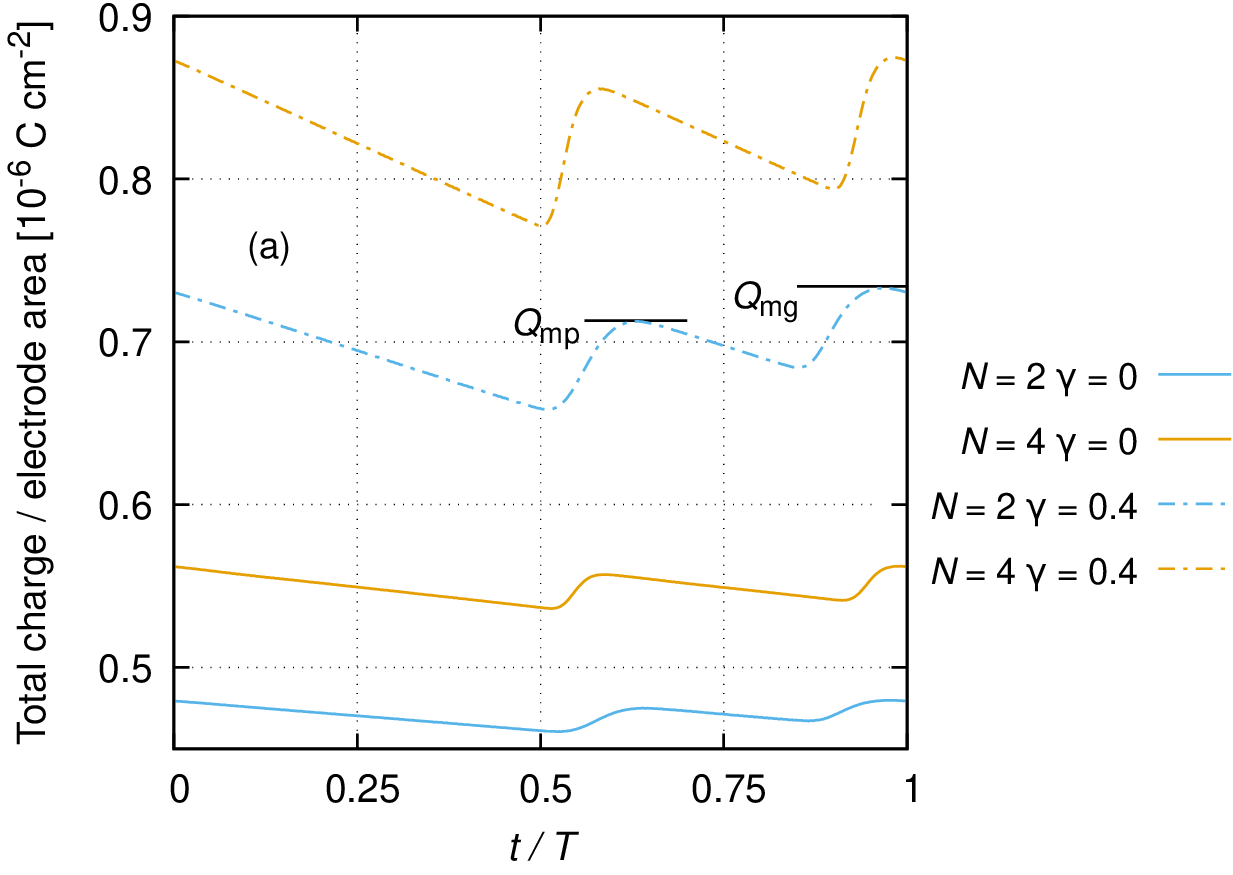}\\
\includegraphics[width=0.48\textwidth]{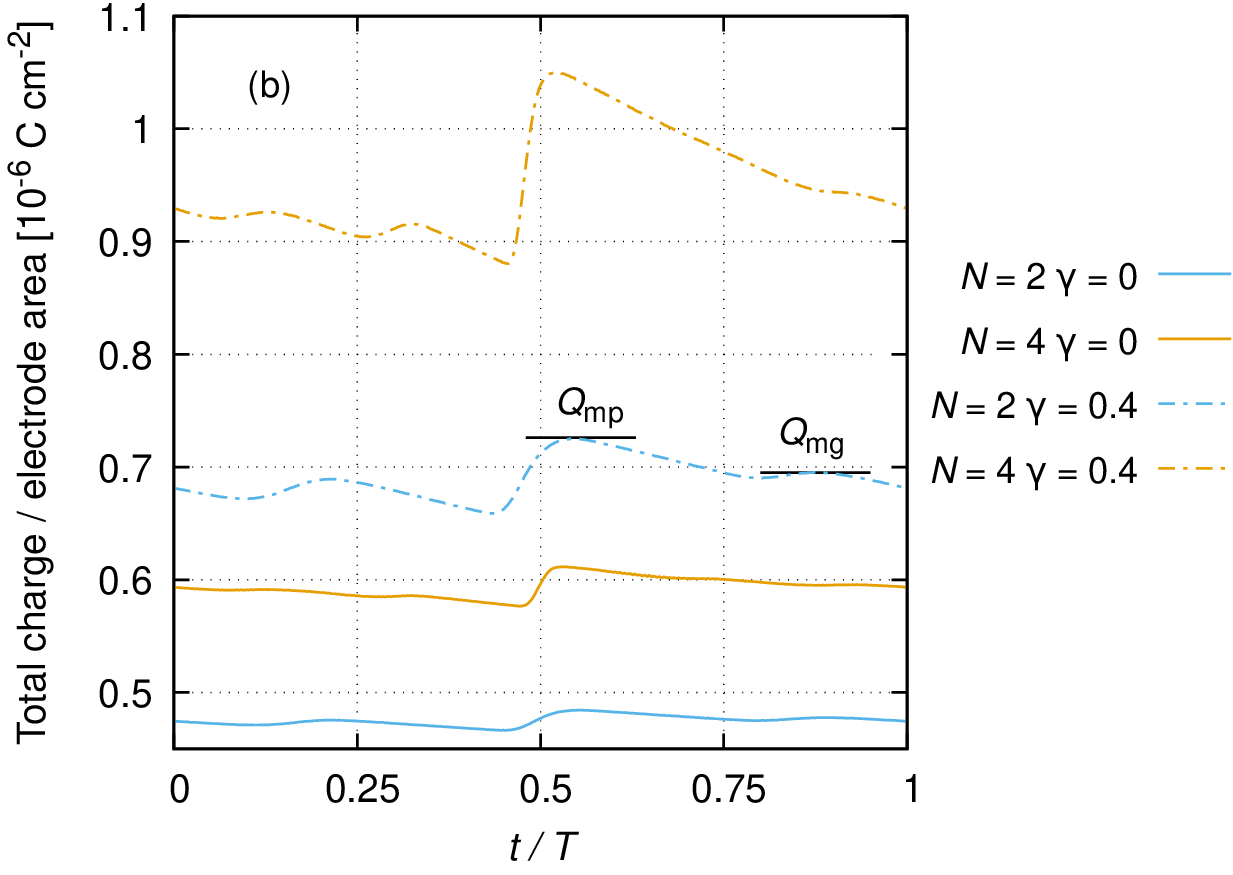}
\caption{Total uncompensated charge per unit area (of 1 cm$^2$) as a function of time within a fundamental RF period for (a)  $\theta=90^\circ$ and (b)  $\theta=180^\circ$, for various values of $N$ and $\gamma$. Parts of the curves with a slow decay correspond to the continuous losses of positive ions, while parts with a steep rise correspond to sheath collapses when electrons are lost to the electrodes. The maximum charges in the two sheaths are illustrated for one of the cases in both panels. For more explanation see text. Discharge conditions: Ar at $p$ = 10 Pa, $L$ = 2.5 cm, $f_1$ = 13.56 MHz, $\phi^\ast$ = 300 V.}
\label{fig:Qtot}
\end{figure}

To understand the behavior of $(Q_{\rm mg} / Q_{\rm mp})^2$, the time dependence of the total charge $Q_{\rm tot}(t)$ in the plasma (per unit area) is plotted in Figure \ref{fig:Qtot}. for various values of $\theta$, $N$, and $\gamma$. All the curves seen in this figure exhibit slowly decaying parts that correspond to the continuous losses of ions at the electrodes and short, rapidly increasing segments where $Q_{\rm tot}$ increases because of the losses of the electrons. Losses of electrons occur during the times of sheath collapse at either side of the plasma. Taking the case of $N=2$ as an example, according to Figure \ref{fig:slope}(a) the grounded sheath collapses at $t_{\rm g}/T \approx 0.58$, whereas the powered sheath collapses at $t_{\rm p}/T \approx 0.9$. At the time of  collapse of the {\it grounded} sheath $Q_{\rm tot}$ resides in the {\it powered} sheath, thus the peak of $Q_{\rm tot}$ at $t_{\rm g}$ in Figure \ref{fig:Qtot}(a) can be associated with $Q_{\rm mp}$ (illustrated for the $N=2$, $\gamma=0.4$ curve). Similarly, $Q_{\rm tot}$ at $t_{\rm p}$ peaks at a value of $Q_{\rm mg}$. 

Having understood this, we can look now at the changes of $Q_{\rm mp}$ and $Q_{\rm mg}$ resulting from an increase of $\gamma$. When $\gamma$ is increased from zero, more ionization occurs near the powered electrode (because of the higher peak sheath voltage there, see above) and, consequently, the ion flux to the powered electrode increases more than the ion flux to the grounded electrode. As the flux of the ions and the electrons to either of the electrodes must compensate each other on time average, an increased ion flux at one electrode also requires an increased electron flux at the same electrode. Recall that the electron flux can flow only during the collapse of the sheath. Due to the higher ionization at the powered side of the plasma the electron flux at the powered electrode will get enhanced more than at the grounded electrode. Consequently, the total charge in the plasma, $Q_{\rm tot}$, will be increased more at the time of sheath collapse at the powered side, as compared to that at the grounded side. According to the explanations about the behavior of $Q_{\rm tot}(t)$ provided above, $Q_{\rm mg}$ will get more enhanced than $Q_{\rm mp}$ when $\gamma$ is increased, leading to the increase of  $(Q_{\rm mg} / Q_{\rm mp})^2$, as observed. 

The scenario at the higher number of harmonics, $N=4$, is somewhat different. In this case, a significantly faster sheath expansion is observed in Figure \ref{fig:slope}(a) after the sheath collapse at the powered side of the discharge as compared to the $N=2$ case (compare the falling slopes of the red solid vs. chain lines, around $t/T=0$). This fast sheath expansion, in contrast with the $N=2$ case, gives rise to a nearly symmetrical  ionization function, which is a result of a compensation between the faster sheath expansion at the powered side and a higher sheath voltage at the grounded side (see Table \ref{table:crossing90}). Therefore, in the $N=4$ case the arguments presented above do not apply for the changes of $Q_{\rm mg}$ and $Q_{\rm mp}$ with $\gamma$. Moreover, the explanation presented above clearly assumed a certain degree of locality of the electron kinetics, i.e., fluxes at a given electrode were assumed to be largely determined by the ionization rate near the same electrode. However, at the conditions considered, energy gain of the electrons at one side of the discharge can contribute to the ionization and particle fluxes at the other side of the discharge as well. This property of the discharge is obviously confirmed by the ("flat top") shape of the ionization source functions plotted in Figure \ref{fig:slope}(b), especially at $\gamma>0$. As a consequence of these effects the small counter-acting changes of the $\overline{n}_{\rm sp} / \overline{n}_{\rm sg}$ and $(Q_{\rm mg} / Q_{\rm mp})^2$ on $\varepsilon$, with increasing $\gamma$ at $N=4$ are difficult to explain. This interplay of these two terms results in a nearly constant $\varepsilon$ and DC self-bias in these cases (see Table \ref{table:crossing90}).

\begin{table}[h!]
\caption{Self-bias voltage ($\eta$) at $\theta=180^\circ$, at $N=2$ and 4, as a function of $\gamma$, as well as the values of the symmetry parameter $\varepsilon$ and the terms involved in  $\varepsilon$.\label{table:crossing180} }
\begin{tabular}{cccccc}
\hline
$N$ & $\gamma$ & $\eta$ [V] & $\overline{n}_{\rm sp} / \overline{n}_{\rm sg}$ & $(Q_{\rm mg} / Q_{\rm mp})^2$ & $\varepsilon$ (Model 3) \\ 
\hline
2 & 0 & 54.60 & 1.274 & 0.851 & 1.084\\
2 & 0.2 & 51.6 & 1.239 & 0.843 & 1.043 \\
2 & 0.4 & 47.7 & 1.194 & 0.825 & 0.985 \\
\hline
4 & 0 & 101.1 & 1.509 & 0.768 & 1.159 \\
4 & 0.2 & 96.8 & 1.472 & 0.755 & 1.111 \\
4 & 0.4 & 84.9 & 1.412 & 0.677 & 0.956 \\
\hline
\end{tabular}
\end{table}


Finally, we address our fourth observation, i.e., the question: why does the self-bias voltage at its extremum (near $\theta=180^\circ$) decrease with increasing $\gamma$ (see Figure \ref{fig:bias_details_pic}(b))? Corresponding data for the DC self-bias voltage, the symmetry parameter and its relevant terms for various values of $\gamma$ are given in Table \ref{table:crossing180}. At $\gamma=0$, ionization is maintained by electron energy gain at the phase of sheath expansion. At $\theta=180^\circ$ (see the thick black lines in Figure \ref{fig:waveforms}), the expansion of the sheath is much faster at the powered electrode and therefore the ionization rate is also higher at that side of the discharge. This results in the high $\overline{n}_{\rm sp} / \overline{n}_{\rm sg}$ values seen in Table \ref{table:crossing180} for both $N=2$ and $N=4$. (Note that the value is higher for the higher $N$ due to the faster sheath expansion induced by the steeper voltage waveform.) As a consequence of this a large $\eta$ is created.

When $\gamma$ is increased from zero, ionization by the secondary electrons starts to play a role. For this contribution the {\it duration} of the expanded phase of the sheaths is a key parameter. For a longer period of expansion, that is actually found at the grounded side of the discharge due to the specific applied waveform, a higher enhancement of the ionization rate is expected. In accordance with this, $\overline{n}_{\rm sp} / \overline{n}_{\rm sg}$ decreases with increasing $\gamma$. 

Using again the argument that a higher ionization rate at the grounded side of the discharge leads to an increase of the total uncompensated charge contained within powered sheath upon the collapse of the grounded sheath, the higher increase of $Q_{\rm mp}$ with respect to that of $Q_{\rm mg}$ (confirmed in Figure \ref{fig:Qtot}(b)), and the consequent  decrease of $(Q_{\rm mg} / Q_{\rm mp})^2$ can be understood. 

As both terms contributing to the symmetry parameter decrease, a strong suppression of the self-bias voltage appears. At the pressure and electrode gap values considered here, energetic electrons created at one side of the plasma also contribute to ionization at the other side of the plasma, as confirmed by the shape of the ionization source functions shown in Figure \ref{fig:slope}(b). Therefore, the discharge has a tendency to become symmetrical at high $\gamma$ values.

\section{Summary}
\label{sec:summary}

In this work, we have examined the establishment of a discharge asymmetry and the concomitant formation of a DC self-bias voltage in capacitively coupled RF discharges driven by multi-frequency voltage waveforms. Computations have been carried out with various values of the coefficients that characterize the electrode surfaces, i.e., (i) the coefficient of elastic electron reflection and (ii) the ion-induced secondary electron yield. The latter ranged between zero and a high value of $\gamma=0.4$, which can characterize high-electron-yield dielectric surfaces. 

The understanding of the computational results has been aided by an analytical model that is based on a voltage balance of the RF discharge. We have shown that this model, in its more complete form when it also includes the charge dynamics (by accounting for the ratios of the net charges in the two sheaths), is able to successfully reproduce and explain the behavior of the DC self-bias voltage as a function of the phase angle between the harmonics of the driving voltage waveform.

The investigations of the surface coefficients indicated that the elastic reflection of the electrons, as long as equal values are used at both electrodes, has a minor influence on the discharge asymmetry and the self-bias voltage. The secondary electron emission coefficient (for which also the same values were adopted for both electrodes) was found to influence the discharge asymmetry and the self-bias voltage in a complicated manner, depending on the phase angle and/or the number of harmonics ($N$). These effects were understood based on the differences of the maximum sheath voltages and the durations of the expanded phases of the sheaths at the two sides of the discharge, as well as on the charge dynamics that is influenced by the ion fluxes to the electrodes. 

At our choice of the gas pressure and the electrode gap, the ionization source function was found to be non-local and a high secondary electron yield induced a tendency to restore the symmetry of the discharge at the conditions of the highest amplitude asymmetry (i.e. in the case of peaks- and valleys-type excitation waveforms).
Further studies could examine such effects at conditions of lower and higher pressures and/or electrode gaps when the nonlocality of the ionization could be enhanced or suppressed.

\acknowledgements

This work was supported by the National Office for Research, Development and Innovation (NKFIH) of Hungary via the grant K-134462, by the German Research Foundation in the frame of the project, ``Electron heating in capacitive RF plasmas based on moments of the Boltzmann equation: from fundamental understanding to knowledge based process control'' (No. 428942393), via SFB TR 87 (project C1), by the National Natural Science Foundation of China (Grant No. 12020101005) and by the grant AP09058005 of the Ministry of Education and Science of the Republic of Kazakhstan.

\section*{Data Availability}
The data that support the findings of this study are available from the corresponding author upon reasonable request.

\nocite{*}
\bibliography{bias}

\end{document}